\documentclass{ar-1col-cranmer}
\usepackage{natbib}
\usepackage{times}
\jname{Ann.\  Rev.\  Astron.\  Astrophys.}
\jvol{57}
\jyear{2019}
\doi{10.1146/((article doi tbd))}

\begin{document}

\markboth{Cranmer \& Winebarger}{Properties of the Solar Corona}

\title{The Properties of the Solar Corona and Its Connection
to the Solar Wind}

\author{Steven R. Cranmer$^1$ and Amy R. Winebarger$^2$
\affil{$^1$Department of Astrophysical and Planetary Sciences,
Laboratory for Atmospheric and Space Physics,
University of Colorado, Boulder, CO 80309, USA;
email: steven.cranmer@colorado.edu}
\affil{$^2$NASA Marshall Space Flight Center, ST13,
Huntsville, AL 35812, USA; email: amy.r.winebarger@nasa.gov}}

\begin{abstract}
The corona is a layer of hot plasma that surrounds the Sun, traces out
its complex magnetic field, and ultimately expands into interplanetary
space as the supersonic solar wind.
Although much has been learned in recent decades from advances in
observations, theory, and computer simulations, we still have not
identified definitively the physical processes that heat the corona
and accelerate the solar wind.
In this review, we summarize these recent advances and speculate about
what else is required to finally understand the fundamental physics
of this complex system.  Specifically:
\begin{itemize}
\item
We discuss recent sub-arcsecond observations of the corona, some of
\\
which appear to provide evidence for tangled and braided magnetic
\\
fields, and some of which do not.
\item
We review results from three-dimensional numerical simulations that,
\\
despite limitations in dynamic range, reliably contain sufficient heating
\\
to produce and maintain the corona.
\item
We provide a new tabulation of scaling relations for a number of
\\
proposed coronal heating theories that involve waves, turbulence,
\\
braiding, nanoflares, and helicity conservation.
\end{itemize}
An understanding of these processes is important not only for improving
our ability to forecast hazardous space-weather events, but also
for establishing a baseline of knowledge about a well-resolved star
that is relevant to other astrophysical systems.
\end{abstract}

\begin{keywords}
heliosphere,
magnetohydrodynamics,
plasma physics,
solar corona,
solar wind,
stellar atmospheres
\end{keywords}

\maketitle
\tableofcontents

\section{INTRODUCTION}
\label{sec:intro}

\begin{figure}[!t]
\includegraphics[width=4.90in]{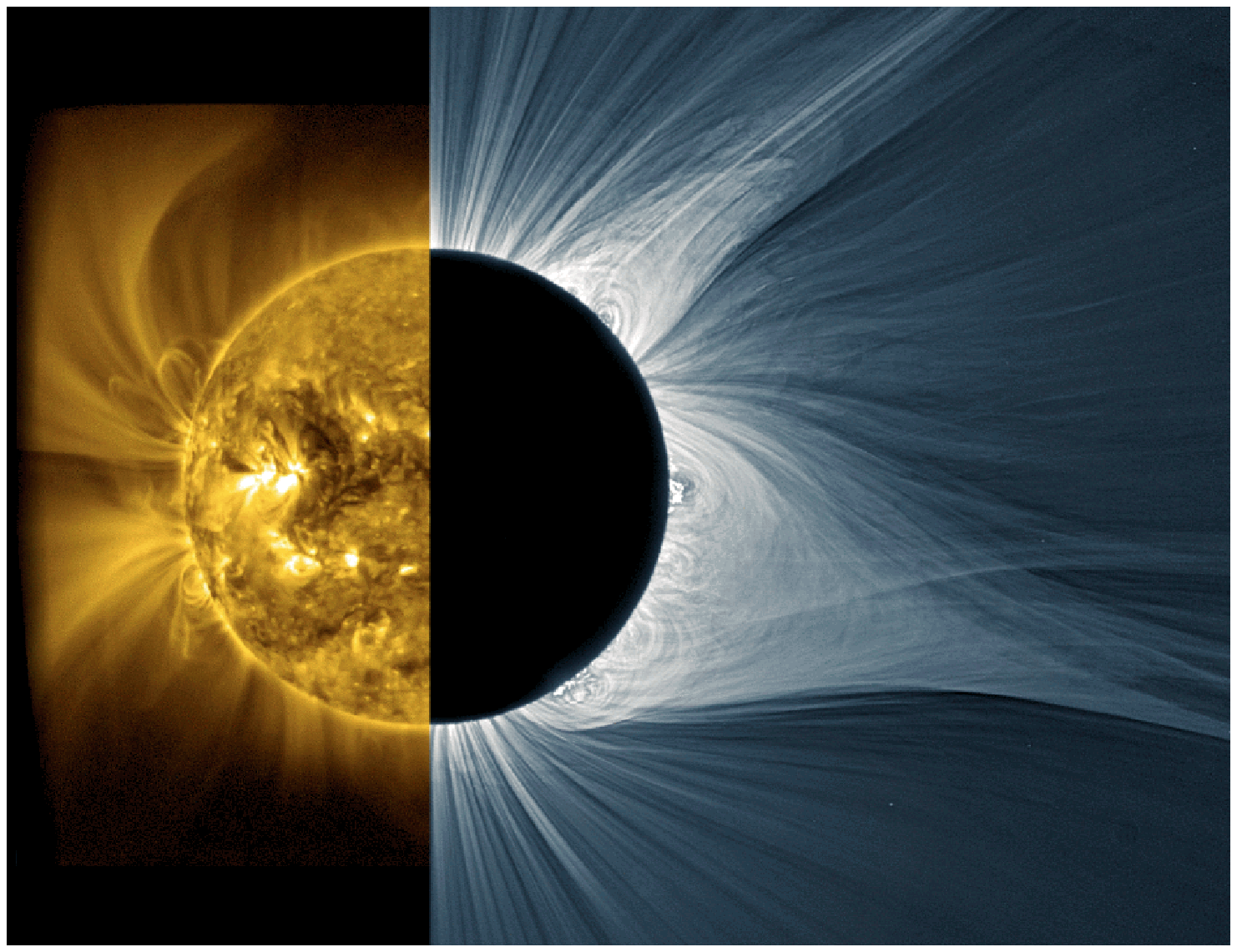}
\caption{Complementary views of the solar corona.
\textit{Left:} Extreme-ultraviolet emission image from the
SWAP (Sun Watcher using Active-pixel-system detector and image
Processing) telescope on the {\em PROBA2} spacecraft \citep{SWAP}.
The image was taken in the 17.1~nm wavelength band on
2014 July 25.
\textit{Right:} Visible-band scattered light from the total
solar eclipse of 2017 August 21, adapted from original images
obtained and processed by M.\  Druckm\"{u}ller, P.\  Aniol, and
S.\  Habbal \citep[see also][]{Dr06}.}
\label{fig01}
\end{figure}

The solar corona is the hot and ionized outer atmosphere of the Sun.
Much of the corona's plasma is confined by the solar magnetic field in
the form of closed loops and twisted arcade-like structures.
In addition, some coronal plasma expands into interplanetary space as
a supersonic outflow known as the solar wind.
\textbf{Figure \ref{fig01}} shows two different views of the corona
and its collection of closed and open magnetic fields.
Despite almost a century of study, the physical processes
responsible for heating the corona and accelerating the solar wind
are not yet understood at a fundamental level.
However, an incredible amount has been learned about this complex
system from continuous advances in observations, theory, and
numerical simulations.
The corona and solar wind have been put to use as laboratories for
studying a wide range of processes in plasma physics and
magnetohydrodynamics (MHD), and they provide access to regimes
of parameter space that are often inaccessible to Earth-based
laboratories.%
\begin{marginnote}[]
\entry{MHD}{magnetohydrodynamics}
\end{marginnote}

The ever-changing corona and solar wind can substantially affect
the near-Earth space environment.
For example, the ultraviolet (UV) and X-ray radiative output of the
corona fluctuates by several orders of magnitude---on timescales between
minutes and decades---and this drives large changes in the ionosphere.
When dynamic variability in the solar wind impacts the Earth's
magnetosphere, it can interrupt communications, damage satellites,
disrupt power grids, and threaten the safety of humans in space.
There is an ever-increasing need to understand how this so-called
space-weather activity affects human society and technology, and
to produce more accurate forecasts \citep[see, e.g.,][]{Ko17}.
Such practical advances are made possible only when there is concurrent
research devoted to answering more fundamental questions such as
``what heats the corona?'' and ``what determines the solar wind speed 
for a given magnetic-field configuration?''

This paper reviews our current understanding of the solar corona and
its connection to the solar wind.
We attempt to provide:
(1) a broad reconnaissance of the present state of the field,
(2) a selection of useful pointers into the primary research literature,
and (3) a brief and selective overview of our shared history.
Because one paper cannot exhaustively cover all work done in such a
large field, we also urge readers to fill in the gaps with other reviews.
Useful surveys of contemporary ideas about the solar corona have
been presented by
\citet{Bi66},
\citet{WN77},
\citet{KIS},
\citet{NU90},
\citet{Lo96},
\citet{As06},
\citet{Kl06},
\citet{GP10},
\citet{PD12},
\citet{Re14},
\citet{SW15},
\citet{Ve15},
and
\citet{Ha18}.
General summaries of the problems and controversies regarding the
solar wind have been presented by
\citet{P63},
\citet{Ds67},
\citet{HA70},
\citet{H72},
\citet{Le82},
\citet{Ax99},
\citet{MV07},
\citet{Zu07},
\citet{BC13},
\citet{Ab16},
and
\citet{Cr17}.
We leave the study of the most explosive events---e.g., solar flares
and coronal mass ejections (CMEs)---to other reviews
\citep[see, e.g.,][]{Fl11,Gp16}.%
\begin{marginnote}[]
\entry{CME}{coronal mass ejection}
\end{marginnote}

\section{BRIEF HISTORY}
\label{sec:history}

\subsection{The Million-Degree Corona}
\label{sec:history:corona}

Although descriptions of an ethereal glow surrounding the eclipsed
Sun can be found going back to antiquity, the first usage of the actual
word {\em corona} (meaning a wreath, garland, or crown) for this
phenomenon was probably by Giovanni Cassini.
On the occasion of the May 1706 solar eclipse, Cassini referred to
``une couronne d'une lumi\`{e}re p\^{a}le,''
or a crown of pale light \citep{WS15}.
Significant progress in understanding the Sun's tenuous outer atmosphere
began to accumulate with the development of spectroscopy in the latter
half of the 19th century.
In 1868, Janssen and Lockyer discovered evidence for
a new chemical element (helium) at the solar limb in the form of a
bright 587.6~nm emission line.
Just one year later, during the solar eclipse of August 1869,
Harkness and Young first observed another emission line, at 530.3~nm,
that did not correspond to any known element.
\citet{Lockyer1869}, in the first issue of {\em Nature,}
discussed the ways this observation was
``...{\em{bizarre}} and puzzling to the last degree!''
Speculation that this line implied the existence of another
new element (``coronium'') persisted for decades, and in that time
a few dozen other mysterious coronal lines were found.

Eventually, \citet{Gr39} and \citet{Ed43} utilized insights from
the new theory of quantum mechanics to determine that the coronal
emission lines were associated with unusually high ionization states
of iron, calcium, and nickel.
This is often assumed to be the primary evidence for a hot corona,
but \citet{A41} assembled several other pieces of observational
evidence that all point in this direction.
For example, the white-light continuum spectrum of the low corona
is dominated by Thomson scattering between photospheric photons and
free electrons.
Thus, the radial variation of the white-light intensity is a probe of
the radial variation of electron density.
\citet{A41} found that the measurements of \citet{Baum37} would be
consistent with hydrostatic equilibrium only for electron
temperatures of about $10^6$~K.
In addition, the lack of sharp Fraunhofer absorption lines in the
coronal white-light spectrum pointed to the existence of substantial
Doppler broadening due to random thermal motions of the electrons,
which also requires similar million-degree temperatures
\citep[see also][]{Gr31,Rg17}.
\citet{A41} also made the earliest estimate of the energy flux required
to heat the corona, and his computed value of 0.2 kW~m$^{-2}$ is
consistent with modern calculations (see Section \ref{sec:theory:flow}).

The corona emits most of its radiation in the ultraviolet and X-ray
parts of the spectrum, but these wavelengths are absorbed strongly
by the Earth's atmosphere.
In the early 1940s, the extent of this atmospheric absorption was not
known, and this pushed experimenters---along with spectrometers
sensitive to UV radiation---to mountain peaks in order to attempt
to extend the solar spectrum into the ultraviolet.
In 1946, a team of researchers launched an ultraviolet spectrometer
on a V-2 rocket for the first time, resulting in both extending the
Sun's ultraviolet spectrum to lower wavelengths and opening a door
to space-based observations that are now the cornerstone of our
knowledge of the solar corona \citep{Tou67}.
The initial rocket flights focused on capturing the solar spectrum 
to determine the elemental makeup of the solar corona.
The data were compared to spectra obtained from ground-based
laboratories and theoretical calculations to identify the emitting
elements and ions.
Data taken from different rocket flights were compared to understand 
the variability of the Sun.
Additionally, spectroheliograms (also called overlapograms, i.e.,
spectrally dispersed images of the Sun on which spatial and spectral
information are overlapping) were made with slitless spectrometers.
These images, typically made along with well-isolated, strong, cool
spectral lines to aid in interpretation, complemented ground-based
observations.

In the 1960s, a pinhole camera was launched on a rocket, and it
obtained the first X-ray photograph of the Sun that provided the
first glimpse of the structure of the million-degree corona
\citep{B63}.
This photo revealed that the high-temperature plasma is not evenly
distributed throughout the solar atmosphere, but is instead confined
to localized ``X-ray plages,'' now commonly called active regions.
These short-duration rocket flights drove the desire for continuous
solar observations above the Earth's atmosphere.
Subsequently, NASA launched several {\em Orbiting Solar
Observatories} ({\em{OSO}}--1 through {\em{OSO}}--8)
from 1962 to 1975, which had autonomous UV,
extreme ultraviolet (EUV), and X-ray instruments on board.%
\begin{marginnote}[]
\entry{EUV}{extreme ultraviolet:
wavelengths of 10--100 nm}
\end{marginnote}
The first space station, {\em Skylab,} in operation from 1973 to 1979, 
also served as a solar observatory, allowing the astronauts to 
operate some of the instruments manually.
These early experiments and their discoveries led to modern-day
observatories on satellites, such as the Japanese-led {\em Yohkoh}
(1991--2001) and {\em Hinode} (2006--present), and the NASA-led
{\em Solar and Heliospheric Observatory} ({\em{SOHO;}} 1996--present),
{\em Solar Terrestrial Relations Observatory} ({\em{STEREO;}}
2006--present), and {\em Solar Dynamics Observatory} ({\em{SDO;}}
2010--present), as well as several smaller class missions.
Over time, the instruments on these observatories have improved the
spatial or spectral resolution, wavelength coverage, cadence or
data volume, or had non-traditional orbits.
There also continues to be a rich sounding rocket and balloon
program that serves as a testbed for new instruments and technologies.

Both historical and modern-day data comprise a broad range of
diagnostics that yield a great deal of information about the solar
corona.
Spectroscopic data in UV, EUV, and X-ray wavelengths provide 
information on the distribution of emission as a function of 
temperature, density, and velocity, and on the composition of the 
coronal plasma \citep[see review by][]{DZM18}.
Images of the Sun in broad X-ray passbands or narrow EUV passbands,
facilitated by the development of multilayer coatings, provide the
spatial distribution of the emission and also a rough estimate of
the emission measure distribution as a function of the temperature
of the plasma.
These observations have been compared to photospheric magnetic
field measurements, which are commonly obtained from both
ground-based and space-based observatories.  
One of the most important realizations from this collective data set
is that the X-ray luminosity of active regions is proportional to the
total unsigned photospheric magnetic flux \citep{Fish98}.
This observation was expanded over 12 orders of magnitude
by including quiet Sun regions, active regions, and stellar coronae
\citep{Pev03}.
These relatively simple observations imply that the magnetic field
plays an important role in the heating of the corona.

Early spectroscopic observations revealed that the composition of the
corona did not always match the composition of the underlying
photosphere. 
Instead, the abundances of a few elements sometimes appeared to be 
enhanced, while the abundances of other elements remained closer to 
their photospheric values. 
The enhanced elements, such as iron and silicon, have low values of 
their first ionization potential (FIP), while the non-enhanced
elements, such as oxygen and neon, have high FIP.%
\begin{marginnote}[]
\entry{FIP}{first ionization potential}
\end{marginnote}
The so-called ``FIP bias,'' i.e., the corona-to-photosphere enhancement 
ratio of elements with FIP lower than about 10~eV, is generally found
to be about 2--4, and it depends strongly on the coronal structure
\citep[see reviews by][]{Me85,Fe92,Sy10}.
The fractionation process that creates the FIP effect is likely 
closely related to the mechanism that heats the corona \citep{Lm15}.

Combining spatially resolved data with spectroscopy can provide
information on individual closed coronal structures, the so-called
coronal loops \citep[see][]{Re14}.
When observed in X-rays, the loops were initially found to be long-lived
and to have densities and temperatures consistent with steady,
uniform heating \citep[e.g.,][]{PK95}.
However, this result was challenged by observations made at EUV
wavelengths \citep{KKA10}.
The densities of the loops are as much as three orders of magnitude
larger than predicted by steady heating \citep[e.g.,][]{Wi03a}, and
the observed pressure stratification does not agree with the expected
gravitational scale height \citep[e.g.,][]{ASA01}.
In addition, the temperatures along the loops are more uniform than
predicted by steady heating \citep{Lz99}.
Though the loops appear to be relatively cool \citep{VK12},
the loops' lifetimes are longer than expected for models of
radiative and conductive cooling \citep{Wi03b}.
Finally, many loops exhibit bulk flows \citep[e.g.,][]{Wi02}
and values of the nonthermal velocity \citep[e.g.,][]{BW16}
that do not appear to match what is expected for several simple
models of uniform heating.
We discuss some of the physical processes underlying these phenomena in
Section \ref{sec:theory:plasma}.

\subsection{The Supersonic Solar Wind}
\label{sec:history:wind}

Starting in the late 19th century, there arose speculation about
a direct connection between phenomena occurring on the Sun
and specific kinds of events taking place on Earth.
\citet{Carr1859} and others took note of the fact that the solar flare
observed in September 1859 was soon followed by strong geomagnetic 
storms (i.e., fluctuations in the Earth's magnetic field) and 
bursts of electric current along telegraph lines.
\citet{Bir08}, reporting on many years worth of data collected on polar
expeditions, made a case that both geomagnetic storms and intense
auroral activity ``...should be regarded as manifestations of an
unknown cosmic agent of solar origin.''
It took several more decades to narrow down the precise physical
nature of this chain of cause and effect.
\citet{Ch18} suggested that the Sun ejects sporadic clouds or beams 
of charged particles into otherwise empty space, and \citet{Hu37}
focused more on ultraviolet radiation as an excitation mechanism
for geomagnetic storms and the aurora.
\citet{Bie51} concluded from the observed properties of comet ion
tails that the solar system appears to be filled with charged
particles (i.e., ``corpuscular radiation'') that are always flowing
out radially from the Sun.

\citet{P58} juxtaposed Biermann's idea of a continuous outflow of
solar particles with the earlier discovery of the high-temperature corona,
and he concluded that these two concepts are inextricably connected.
The high gas-pressure gradient in a hot corona produces an outward
force that counteracts gravity and allows for a time-steady accelerating
flow of plasma away from the Sun.
Parker coined the term {\em solar wind} for this flow, which starts out
slow and subsonic near the solar surface and becomes fast and supersonic
at larger heliocentric distances.
The transition between subsonic and supersonic regimes occurs at a
so-called critical point.
Initially, the \citet{P58} wind solution was criticized for being too
finely tuned; i.e., it seemed unlikely that the system would naturally
choose this one critical solution out of an essentially infinite number
of others that do not become supersonic \citep[see, e.g.,][]{Ch61}.
Although observations soon settled the matter in Parker's favor, it
has also been determined that the critical solution is essentially
a stable attractor of this dynamical system, and that all of the other
possible outflow solutions are unstable \citep{Ve94}.

As noted above, the community had to wait only a few years until
the first {\em in~situ} measurements of particles and fields beyond
the Earth's magnetosphere.
\citet{H72}, \citet{Neu97}, and many others have told the story of the
discovery of the continuous and supersonic solar wind at the dawn of
the Space Age.
In the first few years of interplanetary exploration, it was revealed
that the solar wind often undergoes transitions between a
dense and slow state (i.e., speeds of 250--450 km~s$^{-1}$) and
a tenuous and fast state (500--800 km~s$^{-1}$).
Also, the radial magnetic field frequently alternates sign to form
``magnetic sectors'' that recur with the Sun's 27-day rotation.
These largest-scale plasma structures in the solar wind are now
generally believed to be connected to the topology and geometry of
the Sun's complex magnetic field (see Section \ref{sec:heliosphere:map}).
There is also considerable variability on smaller scales,
such as stochastic MHD turbulence \citep{Co68} and coherent
Alfv\'{e}n waves \citep{BD71}.
The first few decades of solar wind exploration saw missions
that explored inside the orbit of Mercury 
({\em Helios;} \citeauthor{SM91} \citeyear{SM91}),
far past Pluto
({\em Voyager;} \citeauthor{Bu96}\  \citeyear{Bu96}),
and out of the ecliptic plane altogether
({\em Ulysses;} \citeauthor{Md01} \citeyear{Md01}).

\section{ADVANCES IN REMOTE-SENSING OBSERVATIONS}
\label{sec:obs}

With the plethora of ever-improving observations of the corona 
available to solar physicists after the advent of rocket- and
satellite-borne observatories, one might wonder why the solar corona 
heating problem is still a problem.
The answer of course is that many coronal heating theories,
discussed in detail in Section \ref{sec:theory}, predict very
similar observational consequences in the regimes where observations
are easiest to make.
This often makes it difficult to use data from current observatories
to discriminate between the different heating theories
\citep[see, for instance,][]{WW04}.
Here we provide some recent observations that push the boundaries
of current instrumentation.  
Not surprisingly, many of these observations originate in suborbital 
instruments.  

\begin{figure}[!p]
\includegraphics[width=4.40in]{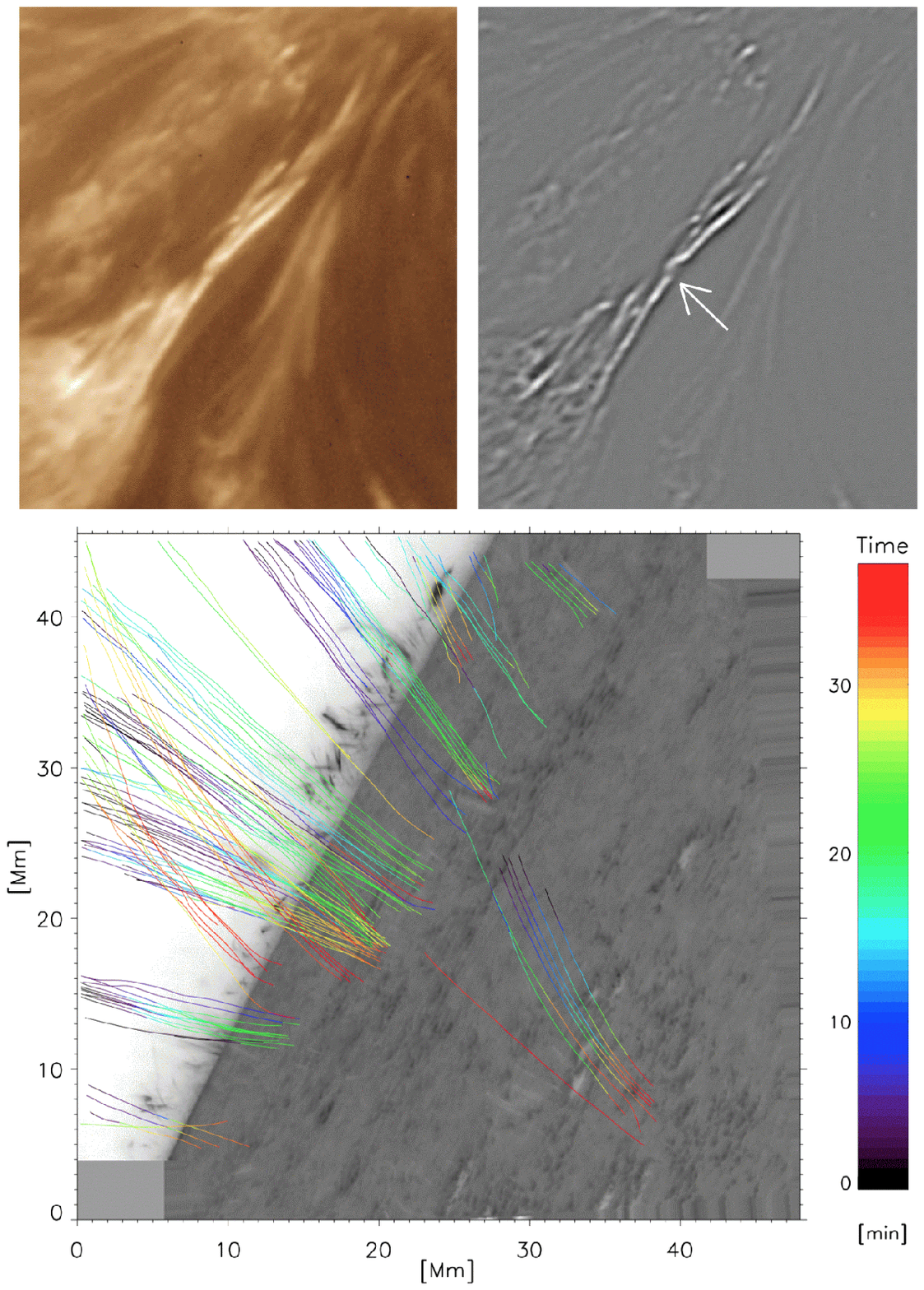}
\caption{Recent observations find conflicting results about
the degree of magnetic tangling of coronal field lines at scales
of order 0.2$''$.
\textit{Top left:} Braided active region structure imaged by
{\em Hi-C} at a wavelength of 19.3 nm, on 2012 July 11.
\textit{Top right:} Enhanced version of the {\em Hi-C} image
made with an unsharp-masking technique \citep[see][]{Ci13};
an arrow highlights a braided structure.  Shortly after this image
was acquired, a heating event was observed at this location
implying that energy stored in the magnetic field was released.
\textit{Bottom:} The paths of coronal-rain plasma parcels, observed
by {\em CRISP} on 2009 May 10, suggest that the field
is smooth and unbraided \citep{AR12}.}
\label{fig02}
\end{figure}

Nearly all coronal heating theories predict that heating will happen 
sporadically on spatial scales much smaller than resolved by current 
instrumentation, meaning that coronal structures or loops that are 
resolved by current instrumentation are actually formed of many 
sub-resolution strands, each tracing a magnetic field line.  
Some theories require that the magnetic strands be twisted or braided,
but normally the resolved coronal structures, such as those shown
in \textbf{Figure \ref{fig01}}, do not show significant evidence
for twisting or braiding on large scales.
Unfortunately, it is unclear whether the strands become more tangled
when observed in higher resolution.
In 2012, the {\em High-Resolution Coronal Imager} ({\em{Hi-C}})
sounding rocket was launched, and it obtained the highest spatial
resolution (0.2--0.3$''$) images of the solar corona in a narrow
EUV wavelength channel.
This data set, for the first time, resolved two examples of coronal
braiding in an active region core by direct observation of the
high-temperature plasma \citep{Ci13}.
Another way of inferring the coronal field structure is by observing
chromospheric plasma at coronal heights in the form of
``coronal rain,'' cool dense plasma that forms high in the solar 
atmosphere and slides down the magnetic field lines.
Coronal rain is thought to be caused by strong, steady heating near
the footpoints of the loops, which gives rise to thermal nonequilibrium
conditions near the loop apex \citep[e.g.,][]{Mu05}.
Using the {\em Crisp Imaging Spectro-Polarimeter} ({\em{CRISP}})
instrument at the {\em Swedish Solar Telescope,} coronal rain was
observed in the H~I Balmer $\alpha$ line at the diffraction limit
of 0.14$''$ \citep{AR12}.
No evidence of coronal braiding was found as the rain traced the
magnetic field strands as it fell.
\textbf{Figure \ref{fig02}} illustrates these apparently conflicting
results.

Different coronal heating theories predict different frequencies of 
heating on the proposed sub-resolution strands.
Correspondingly, changes in the heating frequency imply different
relative amounts of high-temperature ($T > 3$~MK) emission, and
this further suggests that high-temperature plasma could be a key
discriminator in coronal heating.
Unfortunately, high-temperature plasma---which also tends to have
low emission measure---is particularly difficult to detect with
current satellite instrumentation that is most sensitive to the
brighter 1--3 MK plasma \citep{Wi12}.
The {\em Extreme Ultraviolet Normal Incidence Spectrograph}
({\em{EUNIS}}--13) sounding rocket instrument was successful in 
determining that the Fe~XIX line (which has a peak formation
temperature of 8.9~MK) was pervasive and weak through an active
region.
This provided strong evidence that the heating in the active region
was infrequent, potentially from small-scale magnetic reconnection
events called nanoflares.
The {\em Focusing Optics X-ray Solar Imager} ({\em{FOXSI}}--2) 
sounding rocket flight also detected signatures of hot plasma in two 
localized regions in a solar active region, indicating the possibility
of low-frequency nanoflare heating \citep{Is17}.
However, significant evidence also exists to indicate high-frequency
heating, such as expected for wave dissipation models.
The formation of the above-mentioned coronal rain relies on
near-steady and highly stratified heating.
Such energy deposition would not only drive coronal rain, but would
also cause high-temperature structures to disappear and reappear on
long time scales; such behavior has recently been detected \citep{Fr17}.

Another indirect observation in support of magnetic reconnection is
the impact of nonthermal particles as they spiral down magnetic field 
lines and interact with the denser plasma near the magnetic 
footpoints.
\citet{Ts14} discovered evidence for nonthermal particles in a highly 
localized region in data from the {\em Interface Region Imaging 
Spectrograph} ({\em{IRIS}}), which provides high spatial resolution
and high cadence data of the chromosphere and transition region.
These observations indicate that there may indeed be signatures of
magnetic reconnection in observations of improved temporal and
spatial resolutions.  
The presence of nonthermal particles can also be inferred from radio
noise storms \citep{JS18}.

\section{ADVANCES IN IN-SITU MEASUREMENTS}
\label{sec:insitu}

Recent decades have seen vast improvements in the sensitivity,
accuracy, and cadence of instruments that measure the properties
of particles and electromagnetic fields in space.
These measurements have verified that the solar wind is a natural
continuation of the highly structured, dynamic, and nonequilibrium
corona.
The multiple particle populations in the ionized solar wind (e.g.,
protons, helium nuclei, free electrons, and heavy ions) undergo
infrequent collisions with one another, and thus they tend not to
be in a common state of thermal equilibrium.
The particles often exhibit distinctly different bulk flow speeds,
temperatures, and velocity distribution anisotropies, and these
differences are most pronounced in the lowest-density regions
with the fewest interparticle collisions \citep{Ma06,Ka08,Cr17}.
These differences---usually quantified as a function of the charge
and mass of each type of particle---are also useful diagnostics
of the physical processes responsible for heating the plasma.

Precise measurements of heavy-ion abundances and ionization states
are known to carry information about the physical processes that
affected the ions back in the corona.
These composition signatures are ``frozen in'' near the solar surface
and they remain invariant over much of the wind's journey into
interplanetary space \citep[e.g.,][]{Zu07}.
Above a certain point in the low corona, the ions collide with
virtually no more electrons, so they do not undergo any further
ionization or recombination.
These features are often used to trace wind streams down to
specific coronal magnetic structures, and their ionization states
are indirect measures of the corona's electron temperature.
In addition, some types of slow solar wind are seen to contain
subtle enhancements in the abundances of low-FIP elements, which
is similar to the behavior of some coronal loops (see
Section \ref{sec:history:corona}).
The explanation for why the observed distribution of abundances
in the corona and solar wind departs from the photospheric distribution
is still not yet known \citep[see, however,][]{Lm15,Rea18},
and measurements continue to be refined in order to tighten
constraints on the proposed theories.

The heliospheric measurement of MHD turbulence has also
become more sophisticated in the last few decades.
Combining particle and field data from multiple instruments has led
to at least 9 orders of magnitude of coverage in temporal and spatial
scales \citep{BC13,Ki15}.
In the solar wind, there is continuous activity across
frequencies between $10^{-6}$ and $10^{+3}$~Hz.
This corresponds to spatial eddies flying past the spacecraft with sizes
from several astronomical units (AU) down to a fraction of a kilometer.
The smallest sizes overlap with the proton and electron gyroradii
and inertial lengths, and kinetic departures from ideal MHD are
consistently seen at those scales.
These departures include unequal temperatures for electrons, protons,
and heavier ions, differential flows between these species, and
non-Maxwellian velocity distributions.
The nature of the plasma fluctuations is also being revealed by the
use of formation-flying groups of spacecraft.
When these instruments pass through the same parcel of turbulent plasma
at slightly different times and locations, the signals can be combined
to disambiguate the spatial from the temporal fluctuations;
see, e.g., studies from {\em Cluster} \citep{Go15} and the
{\em Magnetospheric Multiscale Mission} \citep{By18}.
This kind of high-resolution data continues to be analyzed with
a wide range of statistical techniques that probe the intermittency,
anisotropy, and multifractality of solar wind turbulence.
For space plasmas in both the MHD and kinetic physical regimes,
turbulence appears to be fundamentally more complex than the
traditional isotropic turbulence found in incompressible
hydrodynamics \citep[see, e.g.,][]{MV11}.

\section{CORONAL HEATING PHYSICS}
\label{sec:theory}

Although the precise mechanisms heating the corona and solar wind
are not yet understood, the ultimate energy source is generally
understood to be the Sun's roiling convection zone.
The following subsections follow the flow of energy from below the
photosphere up through the extended corona, and they summarize as
many of the proposed heating processes as possible.
It should be noted, however, that the ultimate solution of the
coronal heating problem may not involve one single process that acts
in isolation.
The solar corona/wind system is sufficiently complex that it is
likely that different combinations of multiple processes are heating
the plasma in different regions and at different times.

\subsection{The Overall Flow of Energy}
\label{sec:theory:flow}

In the convection zone, thermal energy is transported up by the
rising of hot parcels of gas and the falling of cooler parcels.
Approaching the solar surface, convection carries nearly all of the
energy that is ultimately released as radiation, so the energy flux is
given by $F_{\odot} \approx \sigma T_{\rm eff}^4$, where $\sigma$ is
the Stefan-Boltzmann constant and $T_{\rm eff} = 5770$~K is the
effective temperature.
In the strongly unstable regions of the subsurface convection zone,
$F_{\odot} \approx 63000$ kW~m$^{-2}$.
However, the photosphere tends to sit several scale heights above
the top of the unstable region, and most of that flux escapes as
radiation.
We observe granulation upflows and downflows in the photosphere, but
the residual kinetic energy flux (i.e., $\rho u^3/2$, where $\rho$
is the mass density and $u$ is the bulk flow speed) is only of
order 500 kW~m$^{-2}$.
This is the source of mechanical energy that is often assumed to be
the available pool for energizing the upper atmosphere.
Of course, this estimate does not distinguish between energy carried
upward by rising granules, that carried downward in the intergranular
lanes, and the energy in horizontal motions.

In addition, in many theories of coronal heating there is only a
small filling factor $f$ of the photospheric surface that is
connected magnetically to the corona.
In that case, the energy available at a point in the corona is
diluted by multiplying the mean photospheric flux by $f$.
The filling factor is essentially the coronal magnetic flux density
divided by the field strength in the small photospheric sources,
which tends to be about 1500~G, or close to the equipartition field
strength (i.e., the field strength at which magnetic pressure
balances gas pressure).
To within an order of magnitude, active regions tend to exhibit
$f \approx 0.1$ and weaker-field regions such as the quiet Sun and
coronal holes tend to exhibit $f \approx 0.01$.
Thus, the available energy flux into those regions is probably 
about 50 and 5 kW~m$^{-2}$, respectively.
\citet{WN77} estimated the magnitudes of energy flux required for
coronal heating in active regions, coronal holes, and the quiet Sun
to be about 10, 0.8, and 0.3 kW~m$^{-2}$, respectively, and this
is consistent with the available diluted fluxes.

The process of coronal heating involves both the large-scale transport
of energy from lower to upper layers and the irreversible conversion
of mechanical kinetic energy into random thermal motions of
the particles.
Intermediate steps---such as the excitation of propagating waves or
temporary storage in non-potential magnetic fields---are often
necessary.
The ultimate conversion to thermal energy tends to be most efficient
when the energy is transferred from large-scale, long-lived structures
to small-scale, bursty, and short-lived structures.
Such a transfer is often triggered by some kind of nonlinearity or
instability in the system, and the rate of heating becomes highly
intermittent.
It often makes more sense to refer to the local volumetric
heating rate $Q$ (i.e., the power delivered per unit volume) rather
than the upward vector energy flux ${\bf F}$.
The heating rate $Q$ is formally defined as as
$| \nabla \cdot {\bf F} |$, but it can be expressed approximately 
as $|{\bf F}|/L$, the energy flux distributed over
a coronal loop of length $L$.

\subsection{Acoustic Waves and the Chromosphere}
\label{sec:theory:acoustic}

Convective upflows and downflows are known to give rise to both
stochastic ``noise'' and to globally resonant
pressure-mode oscillations \citep[see, e.g.,][]{S48,NU90,St04}.
Some fraction of this acoustic wave energy propagates up from the
photosphere to the chromosphere, and the longitudinal
velocity amplitude $v_{\parallel}$ generally increases with
increasing height in order to conserve energy flux.
For motions that survive to the point where the amplitude becomes
of the same order of magnitude as the sound speed $c_s$, an
initially sinusoidal wavetrain will evolve into a sawtooth-like
collection of thin shocks.
This is believed to occur no more than 0.5--1 Mm above the 
photosphere, and at these heights the plasma $\beta$ (i.e., the
ratio of gas pressure to magnetic pressure) is either much larger
than or of order unity.
Therefore, much of the subsequent dissipation and heating due to
these fluctuations is often treated in the hydrodynamic (zero
magnetic field) limit.

In general, acoustic waves can be dissipated by collisional transport
effects (e.g., heat conduction, viscosity, or resistivity),
radiative losses, entropy gain at shock discontinuities, or
kinetic wave-particle interactions.
A representative scaling for the volumetric heating rate can be
given as
\begin{equation}
  Q \, \approx \, {\cal E} \,
  \rho v_{\parallel}^2 c_s / \lambda_{\parallel} \,\,\, ,
\end{equation}
where ${\cal E}$ is a dimensionless efficiency factor and
$\lambda_{\parallel}$ is the wavelength.
\citet{CvB07} discussed the limiting cases of dissipation due to
weak ($v_{\parallel} \ll c_s$) and 
strong ($v_{\parallel} \gg c_s$) shocks, and found
${\cal E} \approx 1.8 v_{\parallel}/c_s$ in the weak limit and
${\cal E} \approx 0.4$ in the strong limit.
In most numerical models, the development of shocks and their
rapid dissipation usually means that $v_{\parallel}$ never
becomes larger than about $c_s$ itself.
In the upper chromosphere and low corona, heat conduction also
becomes a significant source of dissipation, whether
the fluctuations are sinusoidal or shock-like.
For this process, ${\cal E} \approx Pe^{-1}$, where $Pe$ is the
P\'{e}clet number, or the ratio of $c_s \lambda_{\parallel}$ to
the conductive diffusion coefficient.

In the weakly magnetized internetwork regions of the Sun (i.e.,
supergranular cell centers), there is still no agreement about
whether the dissipation of acoustic fluctuations is strong enough
on its own to heat the chromosphere.
Existing observations have sometimes pointed to an affirmative
answer \citep{Cu07,BG10} and sometimes to a negative answer
\citep{Ca07,Bk12}.
Numerical simulations are able to reproduce much of the observed
structure and time-dependent dynamics in the non-magnetic
chromosphere \citep[e.g.,][]{CS97}.
However, many simulations tend to produce a highly intermittent 
state; i.e., hot shocks surrounded by larger regions that may be
too dark and cool to produce the steady emission seen in many
chromospheric spectral lines \citep{Ka12}.
No matter the role of acoustic waves/shocks in the chromospheric
energy budget, it is clear that their dissipation tends not to leave
much power available at larger heights to heat the corona
\citep{AW78,CvB07}.
Thus, in recent years the focus has shifted heavily to magnetic
fields and MHD fluctuations as a primary heating mechanism for
both the chromosphere \citep{Js15} and corona
(Section \ref{sec:theory:MHD}).

\subsection{A Plethora of Proposed MHD Processes}
\label{sec:theory:MHD}

Most of the magnetic field lines that are anchored in the photospheric
granulation ($\beta > 1$) are also connected to the low-density
corona ($\beta < 1$), and the complex interplay between these two
disparate regions is far from understood.
When considering the transport of magnetic energy up from the surface,
the Poynting flux ${\bf S}$ helps to specify how much is available.
The injection of energy via the Poynting flux must be balanced either
by dissipation (i.e., heating) or by a long-term buildup of magnetic
energy in the system.
In ideal MHD, with a vector magnetic field ${\bf B}$ and fluid
velocity ${\bf v}$, the Poynting flux is given by
${\bf S} = {\bf B} \times ({\bf v} \times {\bf B})/4\pi$.
Considering the solar surface as a flat plane, the vertical component is
\begin{equation}
  S_z \, = \, \frac{1}{4\pi} \left[ v_z B_{\perp}^2 -
  ({\bf v}_{\perp} \cdot {\bf B}_{\perp}) B_z \right] \,\, ,
  \label{eq:poynting}
\end{equation}
where $z$ and $\perp$ denote the vertical and horizontal components,
respectively \citep[see][]{We15}.
Although the horizontal component of ${\bf S}$ is sometimes considered
as a source of coronal shear \citep{Kn18}, it is mostly the vertical
component that is believed to supply energy to the corona.
In Equation~\ref{eq:poynting}, the first term in square brackets
corresponds to flux emergence from below the surface \citep{Fi99,CI14}.
The second term corresponds to horizontal jostling of an arbitrarily
inclined field line that passes through the surface.
In regions where new flux is not emerging, the jostling term provides
an energy flux that scales as $S_z \approx \rho V_{\rm A}^2 v_{\perp}$,
where $V_{\rm A} = B/\sqrt{4\pi\rho}$ is the Alfv\'{e}n speed.
Typical properties of the photospheric granulation
($v_{\perp} = 1$ km~s$^{-1}$) and the coronal magnetic field
($B = 50$~G) thus appear to be able to supply
energy fluxes of order 20 kW~m$^{-2}$.

In the remainder of this subsection we summarize many of the
mechanisms that have been proposed for dissipating the available
Poynting flux as heat.
Historically, there have been two major schools of thought that
depend on the relative values of two important timescales.
First, the so-called Alfv\'{e}n travel-time $\tau_{\rm A}$
describes how long it takes a linear perturbation to traverse
a significant distance along the coronal magnetic field.
One can write $\tau_{\rm A} = L / V_{\rm A}$, where $L$ is
either the loop length (for closed magnetic fields) or a
representative solar-wind scale height (for open fields).
Second, the photospheric driving timescale $\tau_{\rm ph}$ is a
characteristic time over which the granular motions can
make major changes in the field at the footpoints.
This quantity is often written as
$\tau_{\rm ph} = \lambda_{\perp} / v_{\perp}$, where
$\lambda_{\perp}$ is a horizontal correlation length for
footpoint driving.

Given the above definitions, we can parameterize the MHD coronal
heating rate $Q$ in terms of the vertical Poynting flux, spread
out over the macroscopic scale length $L$, multiplied by a
still-undetermined efficiency factor,
\begin{equation}
  Q \, \approx \, {\cal E} \, \rho V_{\rm A}^2 v_{\perp} / L
  \,\,\, .
  \label{eq:Qfid}
\end{equation}
\textbf{Table \ref{tab01}} provides a sampling of proposals for how
the efficiency factor ${\cal E}$ depends on dimensionless
ratios such as
\begin{equation}
  \Lambda \, = \, \lambda_{\perp} / L
  \,\,\, , \,\,\,\,\,\,\,
  \Theta \, = \, \tau_{\rm A} / \tau_{\rm ph}  \,\,\, .
\end{equation}
For simplicity's sake, dimensionless numerical factors of order unity
are not included in \textbf{Table \ref{tab01}}, and the expressions
themselves tend to be time averages.
The traditional limit of slowly evolving quasi-static equilibria
(i.e., direct-current, or DC theories) corresponds to
$\Theta \ll 1$, and the limit of rapid footpoint-driving that
produces waves and other propagating fluctuations
(i.e., alternating-current, or AC theories) corresponds to
$\Theta \gg 1$.%
\begin{marginnote}[]
\entry{AC}{alternating current,
\\ $\tau_{\rm A} \gg \tau_{\rm ph}$}
\entry{DC}{direct current,
\\ $\tau_{\rm A} \ll \tau_{\rm ph}$}
\end{marginnote}

\begin{table}
\tabcolsep7.5pt
\caption{MHD coronal heating theories: Efficiency scalings 
relative to the Poynting flux}
\label{tab01}
\begin{center}
\begin{tabular}{@{}c|c|c@{}}
\hline
\textbf{Model description} & 
\textbf{Efficiency ($\cal E$)} & \textbf{Example reference} \\

\hline
\multicolumn{3}{c}{Wave Dissipation (AC) Models} \\
\hline

Alfv\'{e}n-wave collisional damping  &
$\Lambda^1 \Theta^2 Re^{-1}$  &
\citet{Os61} \\

Resonant absorption  &
$\Lambda^1 \Theta^1$  &
\citet{Ru97} \\

Phase mixing  &
$\Lambda^1 \Theta^{4/3} Re^{-1/3}$  &
\citet{Rb00} \\

Surface-wave damping &
$\Lambda^{1/2} \Theta^{3/2} (\Sigma / Re)^{1/2}$  &
\citet{Ho85} \\

Fast-mode shock train  &
$\Lambda^2 \Theta^3$  &
\citet{Ho85} \\

Switch-on MHD shock train  &
$\Lambda^3 \Theta^4$  &
\citet{Ho85} \\

\hline
\multicolumn{3}{c}{Turbulence Models} \\
\hline

Kolmogorov-Obukhov cascade  &
$\Lambda^1 \Theta^2$  &
\citet{Ho86} \\

Iroshnikov-Kraichnan cascade  &
$\Lambda^2 \Theta^3$  &
\citet{Ch02} \\

Hybrid triple-correlation cascade  &
$\Lambda^1 \Theta^3 (1 + \Theta)^{-1}$  &
\citet{ZM90} \\

Reflection-driven cascade &
$\Lambda^1 \Theta^2 (f_{+}^2 f_{-} + f_{-}^2 f_{+})$ &
\citet{Hs95} \\

2D boundary-driven cascade  &
$\Lambda^{2/3} \Theta^{1/3}$  &
\citet{HP92} \\

Line-tied reduced MHD cascade &
$\Lambda^1 \Theta^{1/2}$  &
\citet{DG99} \\

\hline
\multicolumn{3}{c}{Footpoint Stressing (DC) Models} \\
\hline

Current-layer random walk  &
$\Lambda^1$  &
\citet{SU81} \\

Current-layer shearing  &
$\Lambda^1 (1 + \Theta^2)^{1/2} (1 + \Lambda^2)^{-1/2} $ &
\citet{GN96} \\

Braided discontinuities  &
$\Lambda^2 \Theta^1$  &
\citet{P83} \\

Flux cancellation &
$\Lambda^1 \Theta^1 (\phi^{8/3} - \phi^{4/3})$  &
\citet{Pr18} \\

\hline
\multicolumn{3}{c}{Taylor Relaxation Models} \\
\hline

Tearing-mode reconnection  &
$\Lambda^1 \Theta^{1} (1 - \alpha L)^{-5/2}$  &
\citet{BP86} \\

Hyperdiffusive reconnection  &
$\Lambda^1 \Theta^{-1} (\alpha L)^2$  &
\citet{vBC08} \\

Non-ideal/slipping reconnection  &
$\Theta^{-1} (\alpha L)^1$  &
\citet{Ya18} \\

\hline
\end{tabular}
\end{center}
\end{table}

The following subsections describe four classes of proposed models in
more detail, but it is worthwhile to first give representative values
for some of the parameters.
For photospheric granulation, $\lambda_{\perp}$ probably ranges
between 0.1 and 1 Mm, and for typical coronal loops,
$L \approx 5$--500~Mm.
Thus, nearly all coronal regions tend to exhibit $\Lambda \ll 1$.
Values for the timescales are more dependent on context.
A typical granule lifetime of 5--10 minutes may be
used for $\tau_{\rm ph}$, but small internal motions inside
intergranular flux tubes may remain coherent for 1 minute or
less \citep{vB11}.
The Alfv\'{e}n travel-time $\tau_{\rm A}$ may be as small as
10 seconds if only the coronal part of the loop is considered, but
it may be longer than 10 minutes if one also counts the travel-time
through both chromospheres at the footpoints \citep{vB14}.
In some theories, the fundamental driving quantity is $v_{\perp}$,
and this varies by more than an order of magnitude depending on
whether it is evaluated at the photosphere ($\sim$1 km~s$^{-1}$)
or in the upper chromosphere (20--40 km~s$^{-1}$).

\subsubsection{Wave Dissipation (AC) Models}
\label{sec:theory:MHD:AC}

Waves are often proposed as an agent for coronal heating because they
provide a way for energy to be generated at the photosphere and then
be transmitted (with minimal losses) up to the corona, where the
conversion to heat can then occur.
The MHD Alfv\'{e}n wave is the least-damped oscillation mode in the
chromosphere, and it is observed ubiquitously in the solar atmosphere
\citep[e.g.,][]{Tz07,Js15}.
However, there are ongoing debates about whether a more specialized
nomenclature should be used to distinguish between different types
of transverse and incompressible waves \citep{Ma13}.
In any case, for this overall class of wave modes, the jostling term
in the Poynting flux can be expressed generally as
$S_z \approx \rho v_{\perp}^2 V_{\rm A}$, which implies an efficiency
factor of at least ${\cal E} = v_{\perp}/V_{\rm A} = \Lambda\Theta$
in Equation \ref{eq:Qfid}.

The first four mechanisms listed in \textbf{Table \ref{tab01}}
differ in the assumed process of Alfv\'{e}n-wave damping.
Specifically, collisional damping in the corona would be dominated
by proton kinematic viscosity $\nu_p$, and the Reynolds number is
defined here as $Re = \lambda_{\perp} v_{\perp} / \nu_p$.
However, the fact that $Re \gg 1$ implies a negligibly small efficiency.
Resonant absorption and phase mixing both require the presence of
relatively inhomogeneous spatial structures in the corona, and these
tend to be on scales that are still too small to observe directly
\citep[see, e.g.,][]{HG95,MA17}.
Similarly, the mode conversion and ultimate dissipation of so-called
surface waves depends on a nonzero value for
$\Sigma = \Delta V_{\rm A} / V_{\rm A}$, the relative change in
Alfv\'{e}n speed over a horizontal scale of order $\lambda_{\perp}$,
normalized by the mean Alfv\'{e}n speed in that region.

\subsubsection{Turbulence Models}
\label{sec:theory:MHD:turb}

In many space plasma environments, conditions are ripe for the
development of a spontaneous and stochastic cascade of
energy from large to small eddies.
This kind of nonlinear turbulent cascade may be present already in
the photosphere \citep{Pt01} and chromosphere \citep{Re08}, and it
is certainly present and strong in the {\em in~situ} solar wind.
Coronal MHD turbulence is likely to develop structures with timescales
bridging the gap between the classical AC and DC limits
($\Theta \approx 1$; see, e.g., \citeauthor{Mi97}\  \citeyear{Mi97}).
Analytic cascade models such as Kolmogorov-Obukhov and
Iroshnikov-Kraichnan are described in more detail by \citet{BC13}.
The Kolmogorov-Obukhov expression in \textbf{Table \ref{tab01}}
produces the same heating-rate scaling as the \citet{GS95}
model of strong and anisotropic MHD turbulence.

For imbalanced turbulence (i.e., due to the collisions of
unequal-strength Alfv\'{e}n-wave packets), the heating rate depends
on $f_{\pm}  = Z_{\pm} / \sqrt{Z_{+}^2 + Z_{-}^2}$, where
$Z_{\pm}$ are the \citet{E50} variables specifying the amplitudes
of the counterpropagating fluctuations.
In closed loops, unequal values of $f_{+}$ and $f_{-}$ can occur
due to different levels of driving from the two footpoints.
In open-field regions, waves propagate primarily up from the surface,
but partial reflection may occur due to wavelengths being of the
same order of magnitude as the radial density gradients.
This kind of reflection-driven cascade has been discussed further
by, e.g., \citet{Ve91}, \citet{Mt99}, and \citet{Ch15}.
The boundary-driven and line-tied cascade models in
\textbf{Table \ref{tab01}} may be equally at home in the
footpoint-stressing category (Section \ref{sec:theory:MHD:DC});
see also \citet{Rp08}, whose DC-type turbulence simulations implied a
continuous range of possible scalings of
${\cal E} \approx \Lambda \Theta^{n}$, with $0 < n < 0.5$.

The scaling laws in \textbf{Table \ref{tab01}} tend to specify only
the inertial-range energy fluxes; i.e., the rates at which
energy cascades from large to small MHD scales.
In steady-state, this rate ought to be equal to the rate of
dissipation and heating, but the physical processes that
perform the heating are not described by the scaling laws.
Thus, alongside the largely MHD-focused macroscopic coronal heating
theories there are also multiple efforts devoted to understanding the
microscopic processes of turbulent dissipation
\citep[e.g.,][]{Ma06,Dr09,Cr14,Pa15}.
Proposed dissipation mechanisms include both collisional effects
(heat conduction, viscosity, or resistivity) and collisionless kinetic
effects (Landau damping, ion-cyclotron resonance,
stochastic Fermi acceleration, Debye-scale electrostatic acceleration,
particle pickup at narrow boundaries, and multi-step combinations of
instability-driven wave growth and damping).
In such a system, the heating and dissipation is likely to occur
intermittently; i.e., as an episodic collection of tiny nanoflare-like
bursts of energy \citep{vB11,Ve15}.

\subsubsection{Footpoint Stressing (DC) Models}
\label{sec:theory:MHD:DC}

\citet{P72} proposed that, in the $\Theta \ll 1$ limit, the magnetic
field in the corona becomes tangled and braided by slow footpoint
motions and the magnetic energy is dissipated via many small-scale
reconnection events.
This is essentially the idea behind the DC current-layer random-walk
scaling relation given in \textbf{Table \ref{tab01}}, and it
also gives rise to intermittent nanoflares.
The same basic scaling (${\cal E} \approx \Lambda$) was also derived
by others both analytically and from the output of numerical
simulations \citep[e.g.,][]{vB86,P88,Hx96,Ng12,Rp18}.
This expression is also the small-$\Theta$ and small-$\Lambda$ limit
of the more general expression given by \citet{GN96}.
The \citet{P83} scaling is similar to the standard braiding model,
but it calls out the special role of the velocity at which reconnection
sweeps discontinuities through the system.
The simple assumption that this velocity is equal to $v_{\perp}$
gives back the same efficiency as the standard current-layer
random-walk model.
However, the alternate assumption that reconnection sweeps through at
a velocity that scales with $V_{\rm A}$ gives the \citet{P83}
scaling relation in \textbf{Table \ref{tab01}}.
\citet{Mn00} tabulated an alternate version in which the
reconnection velocity is given by only the horizontal component of
$V_{\rm A}$, and this ends up giving a result equivalent to the
line-tied cascade case of \citet{DG99}.

Recently, \citet{Pr18} proposed a slightly different DC-type model
that relies on the presence of additional magnetic reconnection at
the chromospheric footpoints (i.e., from small-scale parasitic fields
of opposite sign to the dominant loop polarity) to help power the
large-scale heating.
In \textbf{Table \ref{tab01}}, the dimensionless quantity
$\phi = [F / (\pi \lambda_{\perp}^2 B)]^{1/2}$,
where $F$ is the magnetic flux undergoing reconnection at the base of
the loop and $B$ is the overlying field strength.
Thus, $\phi$ is the ratio of an effective horizontal length-scale
for flux cancellation to the standard footpoint-driving length
$\lambda_{\perp}$.
Just like with turbulence, models that rely on magnetic
reconnection often make assumptions about the micro-scale effects
that ultimately produce the heat.
These non-MHD kinetic processes continue to be studied both
analytically and numerically \citep[see, e.g.,][]{DR12,TB15}.

\subsubsection{Taylor Relaxation Models}
\label{sec:theory:MHD:taylor}

Most DC-type models assume a statistical steady state, in which
nonpotential magnetic energy never gets a chance to build up
very much before nanoflares release the energy as heat.
However, the Sun does sometimes produce highly twisted field
lines in filaments, flux ropes, and sigmoid-shaped cores of
active regions.
The magnetic twist in these regions may be considered as a known
reservoir of energy, and it is often parameterized by the
so-called torsion parameter $\alpha$ (i.e.,
$\nabla \times {\bf B} = \alpha {\bf B}$).
\citet{Ta74} determined that the rate at which energy can be
withdrawn from the reservoir is constrained by a requirement to
conserve magnetic helicity.
The efficiency scalings given in \textbf{Table \ref{tab01}}
show how the heating rate typically increases with increasing
twist when energy is extracted in accord with Taylor's
helicity constraint.
Although $\alpha$ is a signed quantity (indicating the handedness
of the twist), we take its absolute value and express it as a
dimensionless winding number $\alpha L$.

In \textbf{Table \ref{tab01}}, the expression given for the
\citet{BP86} tearing-mode model is approximate; it agrees with
their more involved analytic result in the limit of $\alpha L \ll 1$
and it produces the same asymptotic behavior at $\alpha L = 1$.
This result clearly applies only for $\alpha L < 1$, but observations
often show more twist than that.
\citet{Xe17} studied a collection of twisted active-region loops
and found a mean value of $\alpha L \approx 2.3$, with some having
values as high as 4.
However, these values are still probably below the twist threshold
for the MHD kink instability (i.e., upper limits of order $2\pi$ to
$6\pi$; see \citeauthor{HP79} \citeyear{HP79}).
For the \citet{vBC08} model in \textbf{Table \ref{tab01}}, we assume
that $\nabla \alpha \approx \alpha / \lambda_{\perp}$.
For \citet{Ya18}, we make the same assumption as above that reconnection
sweeps through the system with a velocity that scales with $V_{\rm A}$.

\subsection{Multidimensional Simulations}
\label{sec:theory:multiD}

\begin{figure}[!t]
\includegraphics[width=5.00in]{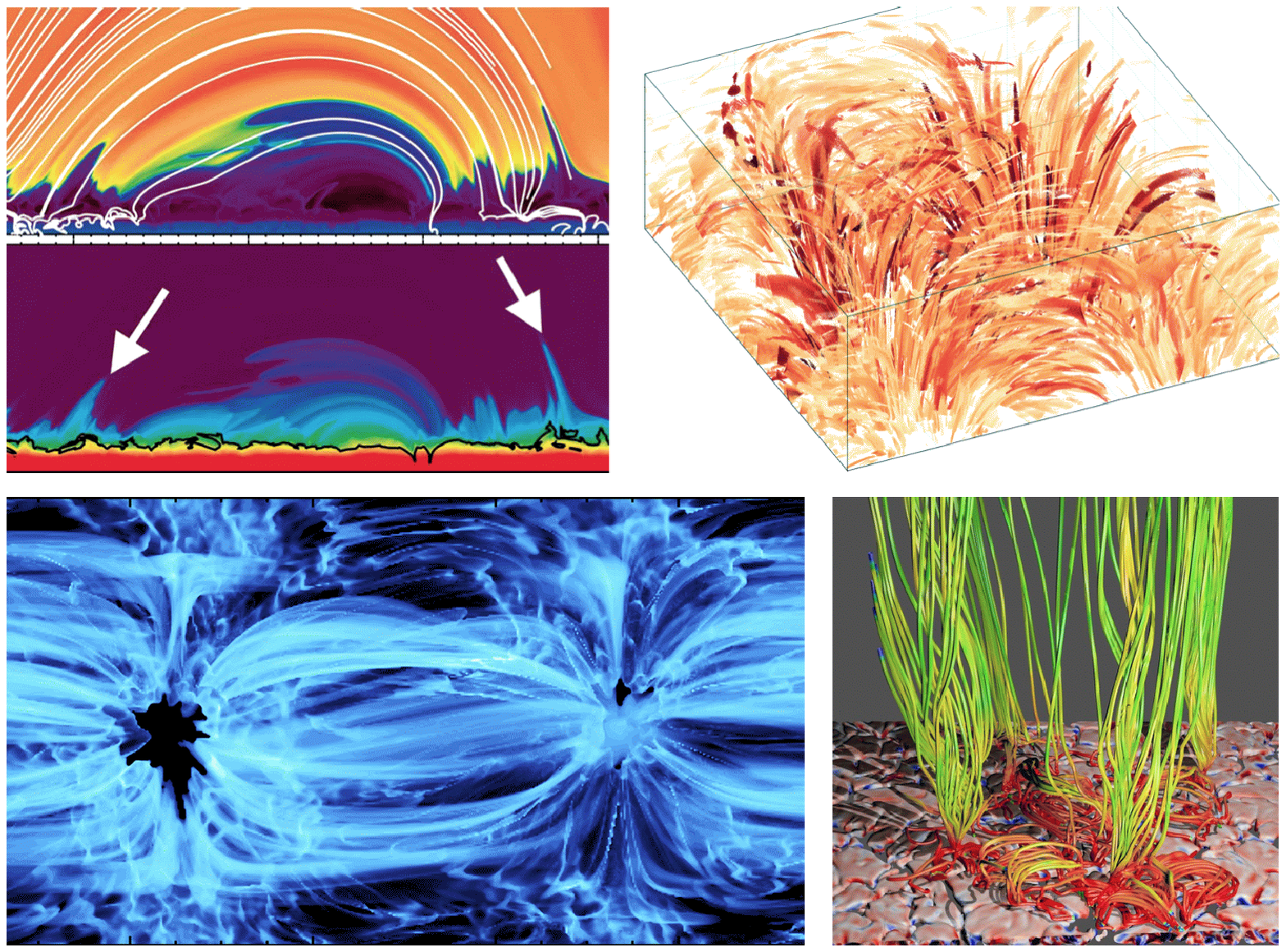}
\caption{A selection of results from multidimensional MHD coronal
models.  Clockwise from upper-left, the illustrated quantities are:
(1) temperature, with magnetic field lines in white (\textit{top}),
and density (\textit{bottom}) from \citet{MS17},
(2) Joule heating rate from \citet{KG17},
(3) impulsively heated magnetic field lines from \citet{Am15},
and (4) synthesized emission measure from \citet{Re17}.
The \citet{Am15} image was published on phys.org on 11 June 2015,
with credit to Tahar Amari (Centre de Physique Th\'{e}orique,
CNRS-Ecole Polytechnique, France).
These simulations are all driven by large-scale surface motions,
such that their heating tends to be dominated by DC-type processes.}
\label{fig0M}
\end{figure}

The analytic scaling relations described above have definite benefits,
but they often fall short of being able to comprehensively explain
a system as complex as the solar corona.
Also, despite the frequent invocation of terms such as intermittency
and nanoflares, these relations tend to be highly averaged in both
space and time.
Thus, the past few decades have seen the development of numerical
simulations that aim to model the fully time-dependent and
three-dimensional structure of the corona
\citep[see, e.g.,][]{We09,Pe15,Dh16}.
\textbf{Figure \ref{fig0M}} illustrates the output from models
constructed by several different groups.
Some of these simulations include self-consistently excited convective
motions in sub-photospheric layers, and others are driven at an
arbitrary lower boundary by parameterized flows that resemble actual
granulation.
These simulations include the conservation of mass, momentum, energy,
and magnetic flux, together with different prescriptions for the
radiation field and the collisional transport coefficients.
They tend to naturally produce a broad range of time/space
intermittency behavior, and a single field line may end up being
heated steadily on one end, and in a bursty manner on its other end
\citep{Pe15}.

At the spatial scales resolved by the current generation of
simulations (e.g., about 0.1 Mm),
there seems to be agreement that DC-type footpoint braiding is the
dominant process, and that it is indeed sufficient to supply the
necessary coronal heating.
However, these coarsely resolved models tend to suppress the generation
of rapid fluctuations such as AC-type waves or MHD turbulence.
Thus, another class of multidimensional numerical experiments has
arisen that aims to follow the internal structure of just one
macroscopic coronal loop, but with greater internal detail
\citep[e.g.,][]{vB11,PC13,Ms18}.
When the photospheric footpoints of these models are driven at
appropriately small space and time scales, they tend to produce
waves and turbulence that dissipate rapidly enough to heat the
corona at reasonable levels.
When these models are driven slowly (i.e., more commensurate with
the DC driving in the coarser simulations), they tend not to produce
much heating \citep{vB14}.
However, the single-loop simulations do not include the effects of
neighboring footpoints that become tangled and twisted up with one
another on larger cross-field scales.
Thus, it is unclear whether a future simulation that resolves both
sets of scales simultaneously will be dominated by AC or DC heating.

\subsection{The Coronal Plasma State}
\label{sec:theory:plasma}

In order to determine which theoretical heating mechanisms apply to
the real solar corona, model predictions must be compared
to observational data.
A variety of approaches has been taken, and some combination of
forward modeling (i.e., taking the model output and synthesizing
artificial observations) and inverse modeling (i.e., processing data
from the telescope to determine the plasma properties in the corona)
must be employed.
A key link in this chain is to understand how a given heating rate
$Q$ gives rise to a known variation of temperature and density
along a magnetic field line.
In general, the corona finds an equilibrium solution that balances
coronal heating with transport and loss terms associated with heat
conduction, radiative emission, and enthalpy transport due to flows.
For coronal loops, these solutions typically have a maximum temperature
$T_{\rm max}$ at the loop apex and a basal gas pressure $P_0$ that
varies slowly along the loop because of the large scale height.
\begin{marginnote}[]
\entry{RTV}{Rosner, Tucker, \& Vaiana}
\end{marginnote}%
The much-cited RTV model \citep{RTV} provides analytic scaling laws
for these quantities under the assumptions of constant $Q$, classical
Spitzer heat conductivity, and a radiative cooling rate that scales
as $\rho^2 T^{-m}$.
The RTV scaling laws are given by
\begin{equation}
   T_{\rm max} \, \propto \, Q^{2/7} L^{4/7} 
   \,\, , \,\,\,\,\,\,\,
   P_{0} \, \propto \, Q^{(11 + 2m)/14} L^{(4+2m)/7} \,\, ,
   \label{eq:rtv}
\end{equation}
and, for simplicity, the normalizing constants in these expressions
are not shown.
Note that a factor of ten change in $Q$ produces a much smaller
(factor of $10^{2/7} \approx 2$) change in $T_{\rm max}$ because
conduction acts as a kind of ``thermostat'' to smooth out the
effects of coronal heating.
Subsequent work has resulted in modified scaling laws that allow for
spatial variability in the pressure and heating rate
\citep[see, e.g.,][]{Se81,AS02,Mt10}.

With the advent of efficient computers, it has become possible to
perform large-scale pixel-by-pixel comparisons between observed
coronal images and synthesized trial images created with a range of
guesses about the heating rate.
Different dependences of $Q$ on quantities such as the coronal field
strength $B$ and the loop length $L$ produce very different patterns
of synthetic EUV and X-ray emission \citep{Mn00,Sj04,Lu08,Fd17}.
For example, \citet{Sj04} found a best fit with observations for
$Q \propto B/L^2$, which is roughly
equivalent to ${\cal E} \approx \Lambda^2 \Theta$, or the prediction
from the \citet{P83} braiding model.
For different data, \citet{Lu08} found a better fit for
$Q \propto B/L$, which comes closer to some of the scalings described
above for Alfv\'{e}n waves (${\cal E} \approx \Lambda \Theta$).
\citet{WW06} pointed out a possible ambiguity between these two
scalings, depending on whether the magnetic field is taken at the
coronal base ($B_0$) or averaged over the loop ($\bar{B}$), because
observations tend to show $\bar{B} \propto B_0/L$.
Of course, the dependence of $Q$ on other parameters besides $B$
and $L$ should not be ignored.
\citet{Tw17} studied the importance of convective suppression in
sunspots to find that bright coronal loops occur when there is at
least one footpoint rooted in the penumbra (high $v_{\perp}$), but
there is virtually no coronal emission when both feet are rooted in
the dark umbra (very low $v_{\perp}$).

\begin{figure}[!t]
\includegraphics[width=6.00in]{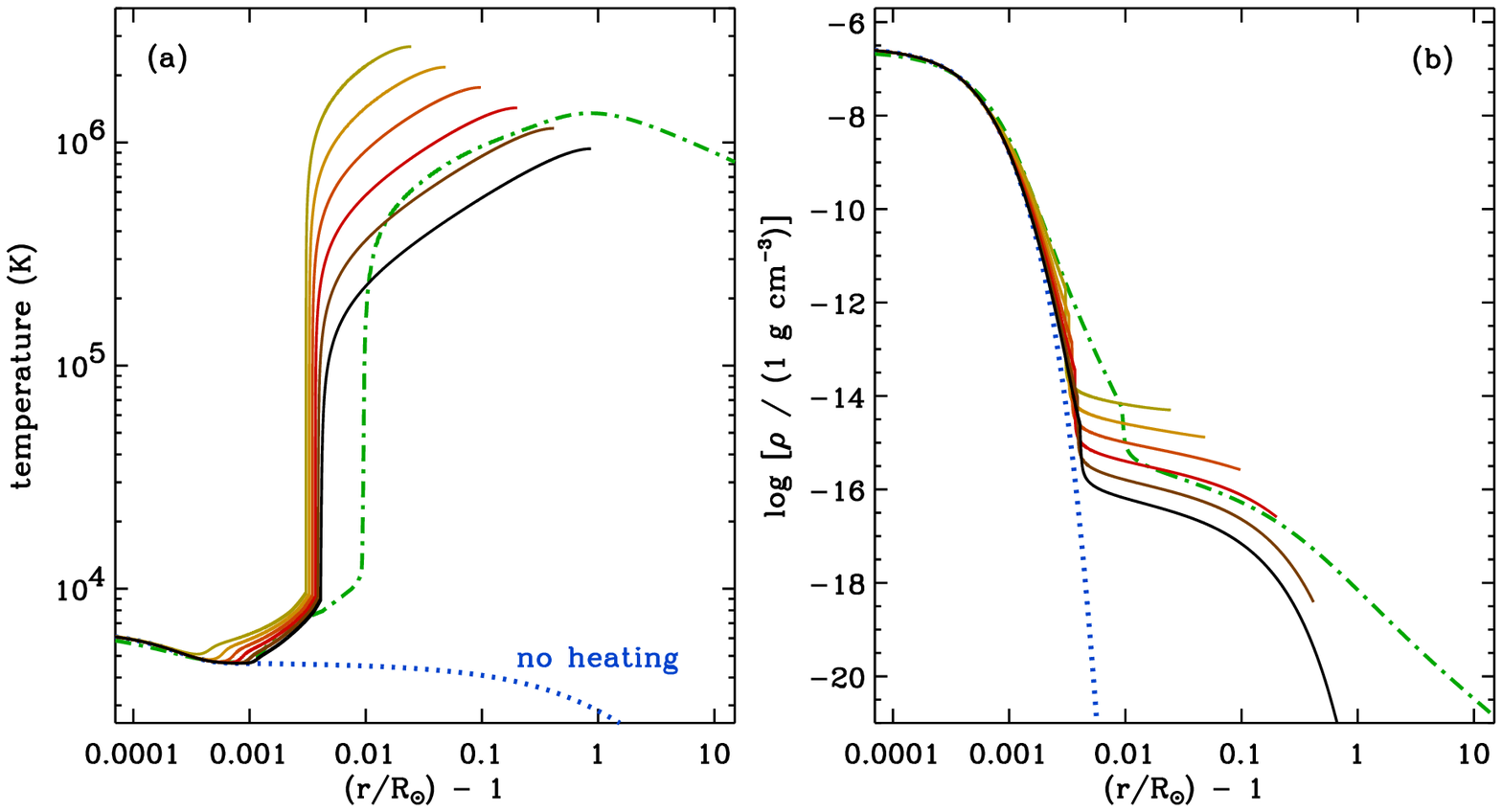}
\caption{Dependence of temperature (\textit{a})
and mass density (\textit{b}) on height above the solar surface
(i.e., with $r = R_{\odot}$ denoting the photosphere, and
heights expressed in units of solar radii) for several
representative models.
Solid curves show closed field-line loops with a range of
lengths $L$ between 30 and 1200 Mm.
The green dot-dashed curve shows an open-field
coronal hole model from \citet{CvB07}.
Another open-field model with no coronal heating whatsoever is also
shown (i.e., pure radiative equilibrium; blue dotted curve).}
\label{fig0T}
\end{figure}

\textbf{Figure \ref{fig0T}} shows example time-steady solutions for
temperature and density for a sequence of closed loops with
different lengths and for an example open-field case \citep{CvB07}.
In the chromosphere ($T < 10^4$~K), the stable solution to the
energy conservation equation comes mainly from a balance between
imposed heating and radiative losses.
However, when the density drops to a point at which radiative cooling
can no longer balance the heating, a rapid transition occurs
to coronal conditions that must also involve heat conduction.
The plotted loop models were meant to emulate empirical
model-atmosphere sequences such as \citet{VAL} and \citet{Fo11}.
Our basic assumption was that $Q \propto L^{-3}$, which is consistent
with a range of heating models from \textbf{Table \ref{tab01}}.
If we also assume $B \propto 1/L$ \citep{JM06},
one can obtain $Q \propto L^{-3}$
from \citet{P83} or \citet{DG99} or just a constant Poynting
efficiency ${\cal E}$.
The loop models in \textbf{Figure \ref{fig0T}} also used $m=0.5$
in Equation~\ref{eq:rtv}, analytic expressions for $T(r)$ in the
corona from \citet{Mt10}, and numerical solutions for the
chromosphere similar to those of \citet{CvB07}.

Computers also allow spatial and time variability in the heating
rate to be incorporated into synthetic observations.
Useful insights have come from zero-dimensional
\citep[e.g.,][]{Kl08}, one-dimensional \citep{Po18}, and
three-dimensional \citep{Mok05} forward modeling with time-
and space-dependent heating.
The multidimensional simulations discussed in
Section \ref{sec:theory:multiD} naturally produce a rapid decline
in the mean heating rate as a function of increasing height, with
$Q \propto \exp(-z/s_{\rm H})$ and $s_{\rm H} \approx 5$--15~Mm
\citep{Pe15}.
Including this effect alone can reproduce many of the observed
properties of the solar corona, such as evolving loops,
larger-than-expected densities, and coronal rain
\citep{Mok16,Wi16}.
Footpoint-stressing models require a finite time between events
to allow for energy to build up in the magnetic field \citep{LFK16}.
Incorporating the time dependence of the heating can also produce
evolving loops and the observed emission measure distribution
\citep{Car15,VD18}.
Finally, the heating rate may be unequal at the two footpoints of a
loop, and this kind of imbalance can drive so-called siphon flows
from one end to the other.
These have been discussed theoretically for many decades, but observed
only rarely \citep[see, e.g.,][]{Hu15}.

Lastly, it is important to note that the coronal plasma state may
not always be described most accurately as a classical MHD fluid.
The creation of a hot corona involves taking some of the cold particles
from below and increasing their most-probable random speeds; i.e.,
broadening their kinetic velocity distributions.
The shapes of these distributions may not always remain Maxwellian,
even in regions where Coulomb collisions are frequent
\citep{MV07,Ec11,Du17}.
It has been proposed that a power-law tail of suprathermal particles
may exist down in the chromosphere, and the fraction of such particles
that escape to larger heights may be enhanced relative to the particles
in the core Maxwellian distribution.
This velocity filtration effect could conceivably produce a hot corona 
without the need for any other heating \citep[see, e.g.,][]{Sc92}.
However, it still requires some mechanism to give thermal energy to
the particles---and thus generate suprathermal tails---in the
chromosphere.
It is unclear what this mechanism could be and how strong it would
have to be to prevent collisions and radiative losses from driving
these particles back into thermal Maxwellian equilibrium.

\section{THE CORONA-HELIOSPHERE CONNECTION}
\label{sec:heliosphere}

\subsection{Physical Processes}
\label{sec:heliosphere:proc}

Parker's (\citeyear{P58}, \citeyear{P63}) original idea of a
gas-pressure-driven outflow is still believed to be responsible
for much of the observed acceleration of the solar wind along open
magnetic field lines.
Thus, the ultimate explanation for the existence of the heliosphere
must come back to an understanding of the coronal heating problem.
In fact, the determination of one key property of the wind---the
total rate of mass loss $\dot{M}$---is so directly related to
coronal heating that it sidesteps Parker's solution of the
momentum equation entirely.
The solar mass-loss rate appears to be set by the same thermal
energy balance that is responsible for setting the base pressure
$P_0$ in coronal loops.
In other words, in both closed and open regions, the dense reservoir
of the chromosphere releases as much plasma as necessary to reach a
time-steady balance between heating, radiative losses, thermal
conduction, and any enthalpy flux due to flows
\citep[e.g.,][]{Hm82,Le82,HL95}.

\begin{figure}[!t]
\includegraphics[width=6.00in]{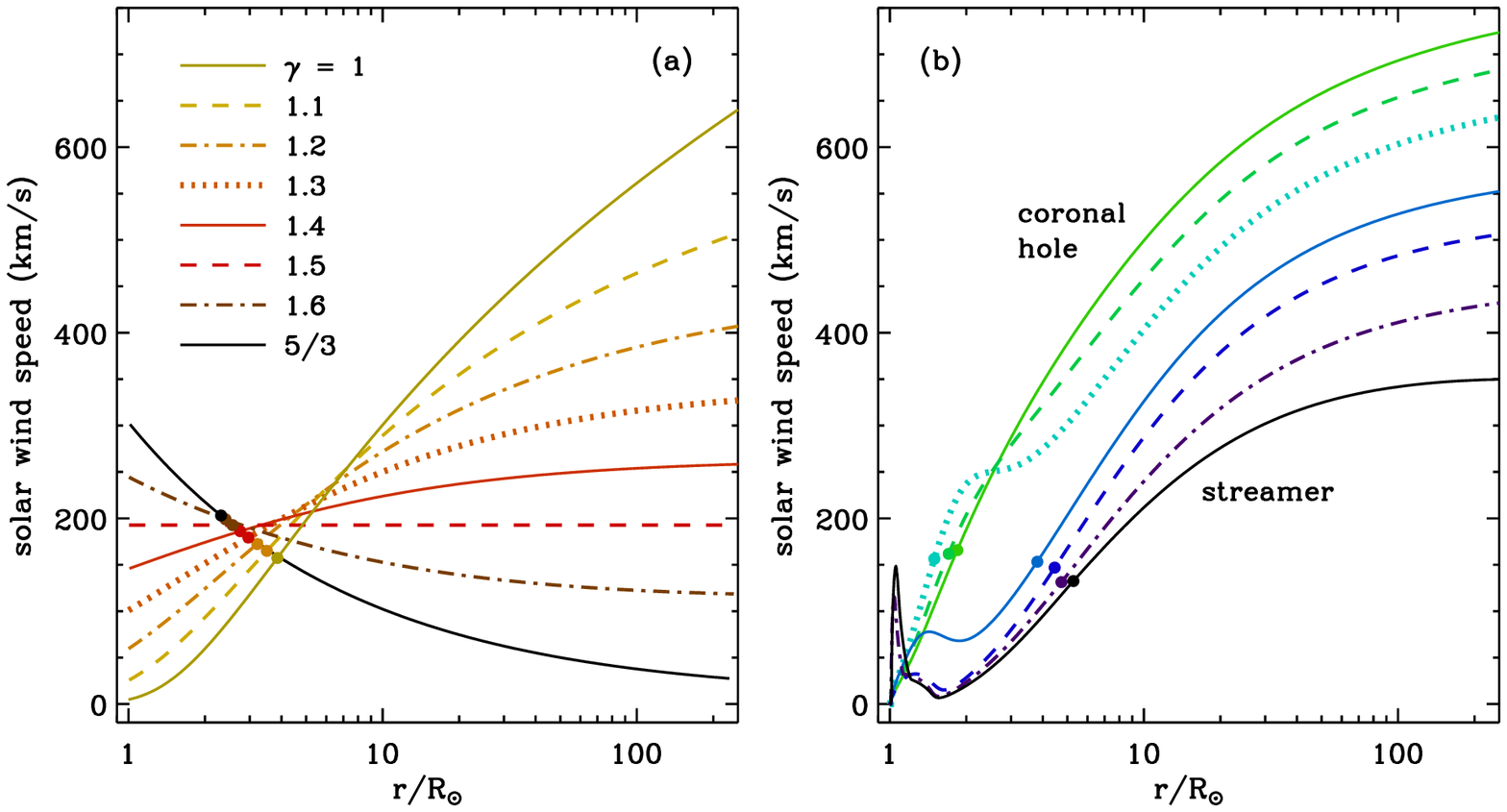}
\caption{Illustrations of one-dimensional models of solar wind
acceleration.
(\textit{a}) Polytropic solutions in spherical geometry, each
computed with $T = 1.5$~MK at the critical point.
(\textit{b}) Self-consistent models of coronal heating via
anisotropic MHD turbulence, for a selection of open field lines
from an axisymmetric solar-minimum magnetic geometry \citep{CvB07}.
In both panels, solid circles denote the locations of Parker's
critical point. Note that 1~AU = 215~$R_{\odot}$.}
\label{fig0W}
\end{figure}

Even without adding any other physical processes besides Parker's
basic gas-pressure gradient, the nature of the acceleration depends
very much on the spatial distribution of coronal heating.
\citet{HA70} and \citet{Ow04} summarized the behavior of solar wind
models with effectively polytropic equations of state; i.e.,
$P \propto \rho^{\gamma}$.
\textbf{Figure \ref{fig0W}a} shows a series of models with a range of
specified $\gamma$ exponents.
The original \citep{P58} isothermal model was equivalent to $\gamma =1$.
The existence of a high temperature that extends to large distances
implies a large gas-pressure gradient that can continue to accelerate
the flow in perpetuity.
Higher values of $\gamma$ imply a more rapidly declining temperature
with increasing distance (i.e., with decreasing density).
\citet{HA70} showed that one requires $\gamma < 1.5$ in order to maintain
an outflow that accelerates through the critical point.
Note that the adiabatic value of $\gamma$ appropriate for a monatomic gas
(i.e., $\gamma = 5/3$) does not allow for an accelerating solar wind.
This means that some kind of non-adiabatic energy addition---either in the
form of extended coronal heating or strong heat conduction---must exist
to prevent adiabatic cooling and to maintain the observed acceleration.

Although there are still gaps in our observational knowledge of the
coronal temperature in regions of solar wind acceleration
\citep[e.g.,][]{Ko06}, we know enough to make the claim that
Parker's gas-pressure gradient must sometimes be supplemented by
other sources of acceleration.
Fast solar wind streams associated with coronal holes have speeds
that ultimately reach 700--900 km~s$^{-1}$ at 1~AU.
This is difficult to explain with observed constraints on gas-pressure
gradients due to the dominant protons, electrons, and alpha particles.
Some have proposed that large-amplitude MHD waves exert enough of a
time-averaged ponderomotive force to provide the extra required
acceleration \citep{AC71,J77}.
There can also be a strong additional outward force due to temperature
anisotropies in the dominant particle velocity distributions.
When the temperature perpendicular to the magnetic field
exceeds that in the direction parallel to the field, there is an
effective magnetic-mirror type force that points in the direction of
weakening magnetic field strength (i.e., outward from the Sun;
see \citeauthor{HI02} \citeyear{HI02}).
Both supplemental sources of acceleration have been proposed
to be present naturally in coronal holes, since these regions tend to
exhibit strong MHD-wave activity and temperature anisotropies
\citep{Ma06}.

Numerical models that account for many of the above processes have been
successful in predicting the observed properties of fast
and slow wind streams \citep[e.g.,][]{Of10,Li14,Go18,Sh18}.
\textbf{Figure \ref{fig0W}b} shows a set of results from \citet{CvB07}
that reproduces the latitudinal variation of solar wind properties at
solar minimum.
This model solves the mass, momentum, and energy conservation equations
along one-dimensional flux tubes of arbitrary geometry using the
reflection-driven cascade model described in
Section \ref{sec:theory:MHD:turb}.
For the highest-speed (open-field polar coronal hole) model, the output
values of temperature and density are shown in \textbf{Figure \ref{fig0T}}.
The slowest-wind models correspond to streamer or cusp geometries that
also have enhanced (i.e., active-region-like) magnetic fields at the base.
In the models, this stronger field produces a small-scale source of
additional time-steady acceleration, which in turn produces a local
maximum in the wind speed of about 100 km~s$^{-1}$ in the low corona.
This may help explain the fan-like outflows of similar magnitude that
have been seen in active regions by {\em Hinode} \citep{Hr08,Bk09,vD12}.

In addition to the wave/turbulence-based models
discussed above, there have been other ideas proposed for the
origin of mass, momentum, and energy in the solar wind.
High-resolution observations of dynamic structures in the chromosphere
(e.g., spicules and jets) show that the solar atmosphere is filled
with rapid, collimated surges of plasma that flow both up and back down.
It has been suggested that a fraction of this plasma becomes heated
to coronal temperatures and thus can be injected directly into the
solar wind \citep[see, e.g.,][]{Mo11,Mc12}.
This scenario is similar to others that emphasize the importance of flux
emergence and interchange reconnection in the supergranular network.
At scales of order 5--30 Mm in the low corona, emerging magnetic
bipoles tend to advect towards the edges of the network and undergo
magnetic reconnection with neighboring flux systems.
This process can transfer hot plasma from closed to open magnetic
field lines and thus drive jet-like pulses of plasma into the
solar wind \citep{Fi99,Ya13}.
Jets are indeed observed both in the chromosphere and corona, but
they tend to be identifiable because they occur intermittently
in time with a small filling factor in volume.
Thus, it is uncertain whether these kinds of impulsive events are
responsible for the majority of the plasma comprising the corona and
solar wind.

\subsection{Mapping and Forecasting}
\label{sec:heliosphere:map}

A long-term objective of solar and heliospheric physics has been to
make accurate predictions of the spatial and temporal distribution of
solar wind plasma properties (usually organized by speed) based on
the state of the corona.
We know of several strong correlations between the coronal magnetic
field and the solar wind at 1~AU, but there are still uncertainties
about the relative contributions of different structures.
The fastest streams (i.e., speeds exceeding 600 km~s$^{-1}$) tend to be
associated with the central regions of large, unipolar coronal holes.
Slow solar wind has been associated with multiple coronal
features---e.g., active regions, helmet streamers,
pseudostreamers, outer boundaries of coronal holes, and
transient jets associated with interchange reconnection
\citep[see][]{Lu02,Bk15,Ab16}---but the relative contributions
from these sources remain difficult to quantify.
\begin{marginnote}[]
\entry{Helmet streamer}{Bipolar magnetic fields stretched out
into the solar wind, with footpoints of opposite polarity}
\entry{Pseudostreamer}{Multipolar magnetic fields stretched
into the solar wind, with footpoints of like polarity}
\end{marginnote}%
There has also been increased interest in how the topological
properties of the Sun's magnetic field may relate to the occurrence
of slow solar wind.
Specifically, \citet{An11} found that there is often a complex
web-like collection of magnetic separatrix surfaces, mainly associated
with pseudostreamers, that corresponds to a 20$^{\circ}$ to 30$^{\circ}$
wide band of slow solar wind around the heliospheric current sheet.
Parcels of slow wind that come from different coronal sources appear
to have different patterns of frozen-in ionization states and FIP
elemental fractionation (see Section \ref{sec:insitu}).
These trends are sometimes used to argue for the prevalance of
magnetic reconnection in the source regions of the slow wind, but
there are also wave/turbulence models that predict similar
patterns \citep[see, e.g.,][]{Cr17}.

Attempts to accurately locate the coronal field lines that connect to
specific fast or slow wind streams are often hampered by the existence
of stochastic processes that can mix and tangle the field lines on
a wide range of scales.
Such processes include small-scale MHD turbulence \citep{Rg09} and
large-scale stream-stream interactions \citep{Ri18}, and their presence
can lead to a loss of information about where on the Sun
a given stream came from.
Because of these ambiguities, there is a lack of agreement about
whether it even makes sense to classify the solar wind into discrete
states or types, and if so, how those classifications should depend
on {\em in~situ} measurements or the successful identification of
coronal source regions \citep[e.g.,][]{Wa09,Ck14,St16,Ne16}.

Despite the difficulties in associating solar wind streams with specific
coronal sources, there is a well-established empirical relationship
between the speed of a parcel of solar wind at 1~AU and the inferred
topological behavior of its approximate magnetic footpoint.
\citet{Lv77} and \citet{WS90} discovered an inverse correlation
between the wind speed and the degree of superradial flux-tube expansion
(i.e., the amount of trumpet-like growth of an area traced out by the
tips of field lines in a compact bundle) between the photosphere and a
reference point at a radial distance of about 2.5~$R_{\odot}$.
This relationship, together with subsequent refinements
\citep[see also][]{AP00,Ri15}, is typically called the WSA model
(after Wang, Sheeley, \& Arge), and it has evolved into an integral
part of modern-day operational space-weather forecasting.
\begin{marginnote}[]
\entry{WSA}{Wang, Sheeley, \& Arge}
\end{marginnote}%
\citet{Kv81} and \citet{WS91} proposed independently that the
physical origin of this effect is related to the existence of
Alfv\'{e}n waves at the coronal base.
Consider a situation in which the upward energy flux of waves is the
same at every point on the solar surface.
A parcel of plasma up in the corona with a low superradial expansion
factor collects wave energy from a larger patch of the surface than
does a parcel of the same size associated with a high superradial
expansion factor.
Thus, the low-expansion regions will have the most vigorous waves
and turbulent fluctuations, the most wave-driven coronal heating,
and thus the most intense solar wind acceleration.
It is probably no coincidence that the central regions of coronal
holes exhibit the lowest superradial expansion factors.

\section{BROADER CONTEXT}
\label{sec:broader}

The Sun is the closest star to the Earth, and for many years it
has served as a template for our understanding of the physical
processes that occur in other stars and even more exotic
astrophysical environments.
This is especially true for the observational signatures of magnetic
fields, hot coronae, and outflowing winds around cool stars, all of
which have been traditionally difficult to detect and characterize
\citep[see, e.g.,][]{Du86,So91,Pg06,Br15,Li17}.
The most luminous stars tend to have quite high rates of mass loss
and dense circumstellar envelopes, so spectroscopic signatures
of their winds are often quite clear.
On the other hand, main-sequence stars similar to the Sun have much
more tenuous winds.
At present, even indirect mass-loss measurements are possible only for
our nearest stellar neighbors, for which interstellar absorption has
not obscured subtle signatures of their astrospheric emission
(i.e., pileup of neutral hydrogen due to winds interacting with the
local interstellar medium; see \citeauthor{Woo18}\  \citeyear{Woo18}).
Also, the growth of techniques such as Zeeman Doppler Imaging
\citep[e.g.,][]{See17} and an increased utilization of high-resolution
X-ray spectroscopy \citep{GN10} have led to much more being known
about the magnetic fields and coronal activity of solar-type stars.

\begin{figure}[!t]
\includegraphics[width=5.00in]{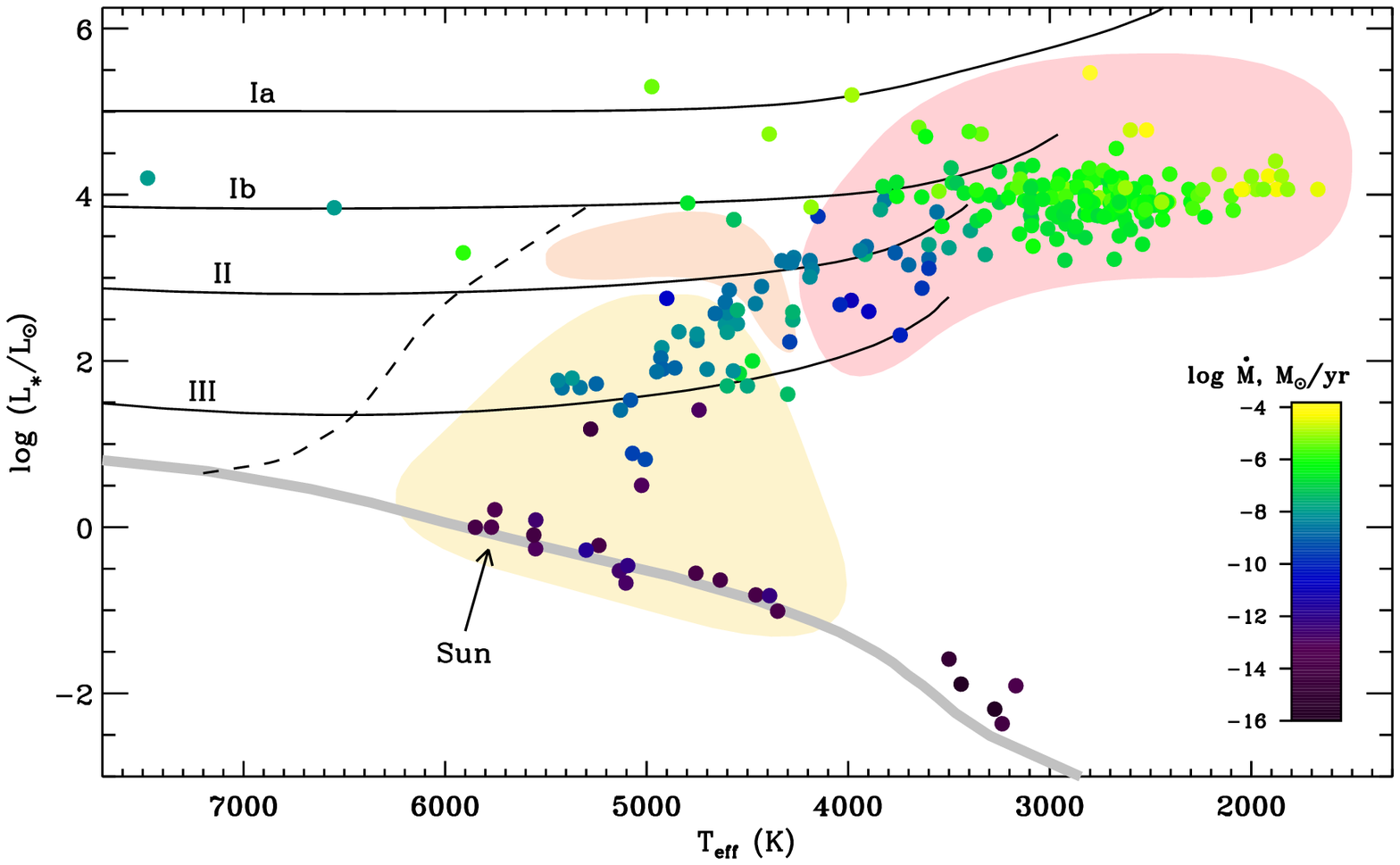}
\caption{Cool-star Hertzsprung-Russell diagram. Symbol colors
correspond to observed mass-loss rates \citep{CS11,Woo18}.
Also shown: zero-age main sequence (\textit{thick gray curve}),
empirical luminosity classes
(\textit{solid black curves;} \citeauthor{dJ87} \citeyear{dJ87}),
and a boundary between stars with and without subsurface
convection (\textit{dashed curve;} \citeauthor{GN89} \citeyear{GN89}).
Larger regions, from lower-left to upper-right, denote
Sun-like stars with hot coronae (\textit{yellow}),
warm/hybrid stars with weak or sporadic coronal signatures
(\textit{orange}), and cool giants/supergiants without
coronae (\textit{red}); see, e.g., \citet{LH79}; \citet{Ay03}.}
\label{fig0H}
\end{figure}

\textbf{Figure \ref{fig0H}} summarizes recent measurements of
mass-loss rates and the regions of parameter space in which coronal
X-rays are typically seen.
The dominant trend appears to be that more luminous stars have
larger mass-loss rates.
This agrees broadly with the idea that each step in the long chain
of processes discussed above---from convective energy transport below
the photosphere to coronal heating above the photosphere---scales with
the total available energy flux flowing through the star
\citep[see also][]{Reim75,SC05}.
The stars most similar to the Sun are those in the lower-left region
of the plot, with high surface gravities and correspondingly small
scale heights.
In the upper atmospheres of these stars, the density drops rapidly
to a point below which radiative cooling cannot balance the heating
(see Section \ref{sec:theory:plasma}) and a million-degree corona
occurs inevitably.
These stars exhibit X-ray and UV emission similar to the Sun's.
However, as one moves to the upper-right part of the plot, the stellar
radii increase, the surface gravities become lower, and thus the
atmospheric scale heights become larger.
Combined with the high rates of mass loss, this leads to high-density
chromospheres that extend for several stellar radii, and there is no
runaway to a hot corona.
For such stars, \citet{Ho83} proposed the existence of cold
wave-driven winds \citep[see also][]{Sz07,CS11}.
There also appears to be a narrow region of hybrid stellar parameters
between the hot and cold domains \citep{LH79}; some these stars
display spectra characteristic of a ``warm'' transition region
\citep{Hy80} and others show UV signatures of weak coronae despite
the lack of X-rays \citep{Ay97,Ay03}.

There are several important avenues of study in astrophysics
and planetary science that depend on (or have developed from)
our understanding of the physical processes that produce the
solar corona and wind.
The following list gives a small and unrepresentative selection.

\begin{enumerate}

\item
In the first few million years after the Sun's formation,
its enhanced mass outflow and UV radiation were probably important
factors in dissipating the primeval atmospheres of the inner planets
\citep{Lam12,Jak18}.
For planets that orbit much closer to their host stars than those in
our solar system, even weak winds (e.g.,
$\dot{M} \approx 10^{-14}$ $M_{\odot}$~yr$^{-1}$)
may have substantial impacts.
The effects of both coronal emission and stellar mass loss need
to be taken into account to accurately determine the age-dependent
masses, densities, and magnetic fields of many types of planets
\citep[e.g.,][]{Hy12,Ga17}.

\item
For young stars, the high-energy coronal radiation responsible
in part for eroding away accretion disks seems to be dominated by
strong stellar flares.
Extrapolating from the present-day Sun, these so-called super-flares
may also be responsible for strong CME-type eruptions of mass and
magnetic flux \citep{Aa12}.
However, there has not yet been a clear and unambiguous detection
of a stellar CME, and it is suspected that strong magnetic
fields may often exert enough of a binding tension force to prevent
eruptive material from escaping \citep{AG18}.
Nevertheless, strong ambient stellar mass loss is observed from
young stars, and the presence of a dense wind can act to shield inner
circumstellar regions from galactic cosmic rays.
This can have a strong impact on the evolution of a star's
protoplanetary disk \citep{Cv15}.

\item
Photospheric elemental abundances of stars can be used as diagnostics
of internal processes such as convective mixing and radiative
acceleration, and they are key to the accurate interpretation of
asteroseismic data \citep[see, e.g.,][]{AP16}.
At layers above the photosphere, cool-star spectroscopy reveals a
diversity of abundance patterns, including sometimes a solar-like
FIP effect (enhanced low-FIP abundances) and sometimes an
inverse-FIP effect (depleted low-FIP abundances).
There is still no complete theory to explain these variations, but
it has been proposed that they arise due to differences in
sunspot/starspot coverage and its impact on wave propagation and mode
conversion in solar/stellar chromospheres \citep{Lm15,DW17}.

\item
Over billions of years, stellar winds return metal-enriched gas
back to the interstellar medium to affect subsequent generations of
stars \citep{Wi00,Da15}.
Fundamental physical processes such as MHD turbulence, magnetic
reconnection, Taylor relaxation, and kinetic particle acceleration are
being invoked with increasing frequency in models of the interstellar
medium \citep{Bu09}, accretion flows around supermassive black holes
\citep{Ro17}, and galaxy clusters \citep{Ba18}.

\end{enumerate}

\section{CONCLUSIONS AND FUTURE PROSPECTS}
\label{sec:conc}

The goal of this paper has been to review some key aspects of
historical and recent advances in our understanding the corona and
solar wind.
Considerable progress has been made over the last few decades in
improving both the quality and quantity of the observational data.
Theoretical models and computer simulations also continue to
proliferate and explain an increasing amount of what we observe.
The solution of the intertwined problems of coronal heating and
solar wind acceleration thus requires us to winnow down the list
of proposed theories and determine which one (or ones) are truly
dominant on the actual Sun.
If these competing ideas could be formalized as a complete set of
mutually exclusive hypotheses, then something like Bayesian reasoning
could be employed to evaluate their relative likelihoods
\citep[see, e.g.,][]{St73}.

Another route toward identifying and characterizing the most important
physical processes for coronal heating is to improve the accuracy
and dynamic range of the simulations.
In Section \ref{sec:theory:multiD} we discussed the goal of
including a broader range of footpoint-driving motions and coronal
field-line interactions in multidimensional MHD simulations.
Only when both AC and DC processes (as well as turbulence and
Taylor relaxation) are allowed to interact with one another without
numerical constraints will the most realistic consequences emerge.
Going beyond single MHD simulations---either into the realm of
nonequilibrium kinetic physics \citep[e.g.,][]{Ce17} or into the
probabilistic arena of large ensembles of simulations \citep{OR17}---is
also becoming possible.
Ultimately, it is crucial to extract from the simulations some key
physical principles that make the results comprehensible to human
beings.
This may take the form of improved coronal-heating scaling laws
\citep[e.g.,][]{Bo16} or, once the simulations reveal which effects
are important and which are ignorable, it may result in completely
new analytic theories.

One of the results of numerical simulations must be to identify
observables that can discriminate between the competing theories
and drive the next generation of solar observations.
Though a wealth of data currently exists for the Sun, multiple
theoretical mechanisms can be shown to be consistent with these
observations.
Undoubtedly the upcoming data from the {\em Parker Solar Probe}
\citep{Fox16} and the {\em Daniel K.\  Inouye Solar Telescope}
\citep{Tri16} will help to differentiate between different mechanisms.
Better measurements of the outer corona (i.e., the extended
acceleration region of the solar wind) are helping to bridge the
gap between the traditionally separate communities of solar and
space physics \citep[see, e.g.,][]{Ko06,Ko08,Df18}.
Improving the spatial resolution of observations and increasing
the temperature sensitivity (especially at higher temperatures)
is undoubtedly important as well.
Migrating successful suborbital instruments (such as {\em{Hi-C,}}
{\em{FOXSI,}} and {\em{EUNIS}}) to orbital platforms---and continuing
to test novel techniques through the sounding rocket program---will
ensure that long-baseline data sets have state-of-the-art
capabilities.

Although the identification of the processes that drive coronal
heating is interesting for its own sake, this problem exists inside
a broader research ecosystem.
\citet{Th14} identified two of the highest-level unsolved
problems in solar physics:
(1) improving the predictability of space weather, and
(2) modeling solar/stellar MHD dynamos from first principles.
Solving the first problem obviously requires improving our
understanding of coronal heating and solar wind acceleration.
Forecasting techniques based on empirical correlations (e.g., the
WSA model) have been successful, but including more of the relevant
physics can only lead to improvements \citep[see, e.g.,][]{Cr17}.
Solving the second problem listed above may also depend on our
knowledge about the mechanisms that produce the corona and heliosphere.
For example, it has been proposed that mass loss in the form of CMEs
must play an important role in a solar-type dynamo by shedding
magnetic helicity that would otherwise build up in the convection
zone \citep{BF00,Bx07}.
In addition, the long-term evolution of a stellar dynamo depends on
the star's rotational history, and that in turn depends on the loss of
mass and angular momentum in the wind \citep{WD67,Bv14,MvS17}.

\section*{DISCLOSURE STATEMENT}

The authors are not aware of any affiliations, memberships, funding,
or financial holdings that might be perceived as affecting the
objectivity of this review. 

\section*{ACKNOWLEDGMENTS}

The authors gratefully acknowledge
Adriaan van Ballegooijen,
Leon Golub,
Stanley Owocki,
Harry Warren,
and Susanna Salom Gay
for many valuable discussions.
S.R.C.\  acknowledges support from
NASA grants {NNX\-15\-AW33G} and {NNX\-16\-AG87G}, NSF grants
1540094 (SHINE) and 1613207 (AAG), and start-up funds from the
Department of Astrophysical and Planetary Sciences at the
University of Colorado Boulder.
SWAP is a project of the Centre Spatial de Liege and the Royal
Observatory of Belgium, funded by the Belgian Federal Science Policy
Office (BELSPO).
This research made extensive use of NASA's Astrophysics Data System (ADS).

\end{document}